# Einstein Chases a Light Beam

Galina Weinstein

Written while I was at The Center for Einstein Studies, Boston University

This is a prelude to a book which I intend to publish. This paper describes my temporary thoughts on Einstein's pathway to the special theory of relativity. See my papers on my thoughts on Einstein's pathway to his general theory of relativity. Never say that you know how Einstein had arrived at his special theory of relativity, even if you read his letters to his wife and friends, and some other primary documents. Einstein gave many talks and wrote pieces, but at the end of the day, he told very little geographical, historical and biographical details pertaining to the years he had spent in the patent office. I thus bring here my jigsaw puzzle and warn the reader again, this is my creation and not Einstein's…

**Introduction**

The most difficult work, with which historians are confronted, is to piece together a coherent jigsaw puzzle from all the scattering and fragmentary pieces of evidence. The great difficulty lies in contradictory evidence. From time to time Einstein answered a few important questions, but his answers are themselves sometimes puzzling or even contradictory. For instance, he could say in 1916 that the Michelson Morley experiment was not a major factor in his development prior to writing his path-breaking papers of 1905.[1] And then some years later, in other circumstances, he would give replies almost just the opposite to this answer. It appears that it is not necessarily because he changed his mind; it is rather the circumstances in which he found himself that dictated an answer and his memory was also a factor.

---

[1] Einstein, Albert (1905a), "Zur Elektrodynamik bewegter Körper, *Annalen der Physik* 17, 1, 1905, pp. 891-921, *The Collected Papers of Albert Einstein. Vol. 2: The Swiss Years: Writings, 1900–1909* (*CPAE*, Vol. 2), Stachel, John, Cassidy, David C., and Schulmann, Robert (eds.), Princeton: Princeton University Press, 1989. (*CPAE* is Collected Papers of Albert Einstein); Einstein, Albert (1905b), "Ist die Trägheit eines Körpers von seinem Energieinhalt abhängig?", *Annalen der Physik* 18, 1905, pp. 639-641 (*CPAE*, Vol. 2, Doc. 24); Einstein Albert, (1905c), "Über einen die Erzeugung und Vervandlung des Lichtes betreffenden heuristischen Gesichtspunkt", *Annalen der Physik* 17, 1905, pp. 132-148 (*CPAE*, Vol. 2, Doc. 14); Einstein, Albert, "Das Prinzip von der Erhaltung der Schwerpunktsbewegung und die Trägheit der Energie", *Annalen der Physik* 20, 1906, pp. 627-633 (*CPAE*, Vol. 2, Doc. 35); Einstein, Albert (1907a), "Über die vom Relativitätsprinzip geforderte Trägheit der Energie", *Annalen der Physik* 23, 1907, pp. 371-384 (*CPAE*, Vol. 2, Doc. 45); Einstein, Albert (1907b), "Über das Relativitätsprinzip und die aus demselben gezogenen Folgerungen", *Jahrbuch der Radioaktivität* 4, 1907, pp. 411-462; 5, 1908, pp. 98-99. (Berichtigungen, errata) (*CPAE*, Vol. 2, Doc. 47; 49). See also: Renn, Jürgen, (ed.) (2005a) *Einstein's Annalen Papers. The Complete Collection 1901-1922*, 2005, Germany: Wiley-VCH Verlag GmbH& Co.

In the crucial years there is a major problem still basically unsolved. The host of evidence and sporadic pieces of primary material do not shed too much light on the course of Einstein's complete way of thinking between 1902 and 1905. Scholars manage to create stories of Einstein's pathway to relativity. However, whatever reading and writing he may have done at this time, Einstein published nothing on the subject of optics and electrodynamics of moving bodies for 3½ years. Surviving correspondence sheds very little light on what happened. There are unfortunately no relevant new letters from this period.[2]

I will try to confront this problem in this paper. All documentary biographies "skip" these 3½ years in Einstein's *scientific life*. They start with Einstein's childhood, they then tell the story of Einstein sitting in the Patent office (and examining patents), Einstein and the Olympian Academy, and then move on to Einstein's later life. However, they say nothing about Einstein's scientific work while at the patent office, because they have no information about it. Einstein wrote his best friend Michele Besso on March 6, 1952 from Princeton about Seelig's 1952 biography:[3] "I already know that this good Seelig is currently dealing with my childhood. This is justified to some extent, since the rest of my existence is known in detail, which is not the case, specifically, concerning the years spent in Switzerland. This gives a misleading impression, as if, so to speak, my life had begun in Berlin!"[4]

Historians of science have been working on this problem and endeavored to present the intricate complicated network that might have led Einstein to the special theory of relativity.[5] I shall start discussing Einstein's pathway to special relativity starting from 1895.

---

[2] Stachel, John, "Einstein and Ether Drift Experiments", *Physics Today* 40, 1987, pp.45-47; reprinted in Stachel, John, *Einstein from 'B' to 'Z'*, 2002, Washington D.C.: Birkhauser, pp. 171-176; p. 175.

[3] Seelig, Carl, *Albert Einstein Und Die Scweiz*, 1952, Zürich: Europa Verlag, or Seelig Carl, *Albert Einstein; eine dokumentarische Biographie*, 1952, Zürich: Europa Verlag.

[4] Einstein to Besso, March 6 1952, letter 182, in Einstein, Albert and Besso, Michele, *Correspondence 1903-1955* translated by Pierre Speziali, 1971, Paris: Hermann. Einstein very likely did not want people to know "in detail" about his "existence" in the Patent Office. Einstein got some of his best ideas in the Patent office, as he later told Besso on December 12, 1919, "I was very interested that you want to go again to the Patent Office, in this weltliche Kloster [worldly cloister] I hatched my most beautiful thoughts, and there we spent such happy days together. Since then, our children have grown and we have grown old boy!" Einstein to Besso, December 12, 1919, Einstein and Besso, Speziali, 1971, Letter 51. Einstein liked describing the Patent Office to his friends as a worldly cloister. Seelig Carl, *Albert Einstein: A documentary biography*, Translated to English by Mervyn Savill 1956, London: Staples Press, p. 56; Seelig Carl, *Albert Einstein; eine dokumentarische Biographie*, 1954, Zürich: Europa Verlag, p. 68.
The worldly cloister appeared to be a private personal cloister.

## 1. Einstein Believes in the Ether

In the spring of 1895 Einstein went to take the entrance examinations for the polytechnic. He failed the humanistic-linguistic part and went to Aarau to finish his secondary education.

In this same year 1895 Lorentz published his seminal work, *Versuch einer Theorie der electrischen und optischen Erscheinungen in bewegten Körpern* which Einstein would later read.[6]

Before Einstein went to Zurich to take the examinations – probably a few weeks or months earlier – he *sent* from Milan to his uncle Cäser Koch an essay, "Über die Untersuchung des Ätherzustandes im magnetischen Felde" – "On the Investigation of the State of the Ether in a Magnetic field". The essay was written in sloping and spidery Gothic script on five pages of lined paper.[7]

---

Einstein wrote Koch, "My dear Uncle, […] I always hesitated to send you this [attached] note because it deals with a very special topic; and besides it is still rather naïve and imperfect, as is to be expected from a young fellow like myself. I shall not mind it at all if you don't read the stuff […]."[8]

Neither letter nor paper were dated; the letter was dated summer 1895 by reference to Einstein's first attempt to enter the ETH, and the paper was dated by the fact of its enclosure in the letter, on the assumption that it was written shortly before the letter.[9]

Gerald Holton wrote, "The covering letter that Einstein had sent with the essay to his uncle, Caesar Koch, received later the added note in Einstein's own hand: '1894 or 95. A. Einstein (date supplied 1950).'"[10] That is, in 1950 Einstein dated the essay to be from 1894 or 1895.

Holton and Abraham Pais analyzed Einstein's essay. Einstein wrote,[11]

"The marvelous experiments of Hertz most ingeniously elucidated the dynamic nature of these phenomena, the propagation in space, as well as the qualitative identity of these motions with light and heat".

Holton explained that the essay "shows that Einstein had already encountered Hertz's work on the electrodynamic field".[12] Pais added, "I do not know how he became aware of Hertz's work. At any rate, it is evident that at that time he already knew that light is an electromagnetic phenomenon but did not yet know Maxwell's papers".[13]

Einstein wrote in the essay,[14]

"The most interesting, and also most subtle, case would be the direct experimental investigation of the magnetic field formed around an electric current, because the exploitation of the elastic state of the ether in this case would permit us a look into the enigmatic nature of electric current. The analogy would also permit us to draw sure conclusions about the state of the ether in the magnetic field surrounding the electric current, provided the previously mentioned investigations attain their ends".

Pais wrote, "The main questions raised in the essay are, how does a magnetic field, generated when a current is turned on, affect the surrounding ether? How, in turn, does this magnetic field affect the current itself?"[15] Holton also says that Einstein "was thinking up experiments to probe the state of the ether which, he said, 'forms a magnetic field' around electric current."[16]

---

[8] Einstein to Caesar Koch, Summer 1895, *CPAE*, vol 1, Doc 6.
[9] *CPAE*, Vol 1, Doc 5 and Doc 6, note 1, p. 6, note 1, 10.
[10] Holton, Gerald, *Thematic Origins of Scientific Thought: Kepler to Einstein*, 1973/1988, Cambridge, Ma: Harvard University Press, pp. 396-397.
[11] *CPAE*, Vol 1, Doc 5.
[12] Holton, 1973/1988, p. 377.
[13] Pais, Abraham, *Subtle is the Lord. The Science and Life of Albert Einstein*, 1982, Oxford: Oxford University Press, p. 131.
[14] *CPAE*, Vol 1, Doc 5.
[15] Pais, 1982, p. 131.

Einstein wrote in the essay,[17]

"First of all, however, it has to be possible to prove that there does exist a passive resistance against the production of the magnetic field by the electric current, and that this [resistance] is proportional to the length of the current circuit and independent of the cross section and material of the conductor".

Pais wrote that, "the young Einstein discovered independently the qualitative properties of self-induction (a term he did not use). It seems clear that he was not yet familiar with earlier work on this phenomenon (by Joseph Henry in 1832).[18]

Holton reminds the reader, "It would be an error to think of that essay in any way as a draft of ideas on which the later relativity theory was directly based, or even to regard it necessarily as his first scientific work". Holton stresses that "what is most significant about" Einstein's first essay from age sixteen "is the idea of the light beam as a probe of a field. From the contemplation of how to measure the wavelengths of such a beam, it would be only a small step to the recognition of the paradox Einstein discovered soon afterwards at the Aarau school".[19]

In addition the essay "On the Investigation of the State of the Ether in a Magnetic Field" could be a very early predecessor to the magnet and conductor thought experiment; but at this stage Einstein believed in the ether.

One can ask with good reason, whether Einstein started to think about the problem which led him to the special theory of relativity already in Pavia in 1894-1895? Was the starting point in Milan or was it only next year in Aarau?

In 1920 Moszkowski seemed to have proposed his answer to this question:[20]

---

[16] Holton, 1973/1988, p. 377.
[17] *CPAE*, Vol 1, Doc 5.
[18] Pais, 1982, p. 131; *CPAE*, Vol 1, note 7, p. 9; Henry, Joseph, "On the Production of Currents and Sparks of Electricity from Magnetism", *The American Journal of Science and Arts* 22, 1832, pp. 403-408.
[19] Holton, 1973/1988, p. 377.
[20] Moszkowski, Alexander, *Einstein the Searcher His Works Explained from Dialogues with Einstein*, 1921, translated by Henry L. Brose, London: Methuen & Go. LTD; appeared in 1970 as: *Conversations with Einstein*, London: Sidgwick & Jackson, 1970, pp. 227-228; Moszkowski, Alexander, *Einstein, Einblicke in seine Gedankenwelt. Gemeinverständliche Betrachtungen über die Relativitätstheorie und ein neues Weltsystem. Entwickelt aus Gesprächen mit Einstein*, 1921, Hamburg: Hoffmann und Campe/ Berlin: F. Fontane & Co, p. 225.
Es ist nämlich Tatsache, daß schon in dem Schüler von Aarau Probleme Wurzel geschlagen hatten, die bereits an der Peripherie der damals möglichen Forschung

"For it is a fact that even in the pupil at Aarau problems had taken root that already lay in the vanguard of research at that time. He was not yet a finder, but what he sought as a sixteen-year-old boy was already stretching into the realms of his later discoveries. We have here simply to register facts, and to abstain from making an analysis of his development, for how are we to trace out the intermediate steps, and to discover the sudden phases of thought that lead a very young Canton pupil to feel his way into a still undiscovered branch of physics? The problem that occupied him was the optics of moving bodies, or, more exactly, the emission of light from bodies that move relatively to the ether. This contains the first flash of the grandiose complex of ideas that was later to lead to a revision of our picture of the world. And if a biographer should state that the first beginnings of the doctrine of relativity occurred at that time, he would not be making an objectively false statement."

## 2 Einstein Chases a Light Beam

Between 1895 and 1896 in Aarau, Einstein was sixteen and the story complicates. If we thought the starting point could be Milan, we better return three or four years back to Munich. Einstein is twelve or thirteen years old. He meets once a week Max Talmud. The latter exposes him to Aaron Bernstein's *Naturwissenschaftliche Volksbücher: Wohlfeile Gesammt-Ausgabe*.[21] Einstein is thrilled and reads Bernstein's books enthusiastically.

In volume sixteen Bernstein describes the wonders of the skies, and then dedicates a chapter to each planet; finally he invites his readers to join him for a fantasy journey into space. Under the title, "Eine Phantasie-Keise im Weltall", "1. Die Abreife", Bernstein described his imaginary journey,[22]

Suppose you want to perform a voyage to space. You need a passing-card, and some provisions, food, a suitcase. Although our voyage is going to be very fast, we are going deep into space. In our suitcase we will take our thoughts. "We travel by water?

---

lagen. Noch war er kein Finder, allein was er als Sechzehnjähriger suchte, ragte schon in die Gebiete seiner späteren Entdeckungen hinein. Hier heißt es: einfach registrieren, mit Verzicht auf die Analyse seines Werdegangs, denn wie sollen wir die Zwischenglieder aufspüren, die Denksprünge, die einen blutjungen Kantonsschüler dahin führen in eine noch gänzlich verschlossene Physik hineinzutasten ? Das Problem, das ihn beschäftigte, betraf die Optik bewegter Körper, genauer: die Lichtaussendung von Körpern, die sich relativ zum Äther bewegen. Darin liegt die Witterung des großen Ideenkomplexes, der weiterhin zur Umgestaltung des Weltbildes führen sollte. Und wenn ein Biograph hinschriebe, daß die Uranfänge der Relativitätslehre bis in jene Zeiten zurückfallen, so würde er nichts objektiv Falsches behaupten.
[21] Bernstein, Aaron, *Naturwissenschaftliche Volksbücher: Wohlfeile Gesammt-Ausgabe*, 1870/1897, Berlin: Ferd. Dümmlers Berlagsbuchhandlung.
[22] Bernstein, 1870/1897, Sechzehnter Teil, p. 54.

On the back of the horse? By train? None of that! We travel with the help of an electrical telegraphical apparatus!"

A few years later, Einstein at school in Aarau imagined a journey on a light beam as well (not exactly on a telegraphic signal); the thought experiment of him chasing a light beam. Friedrich Herneck thought that Bernstein might have inspired Einstein when he propounded his Aarau thought experiment. Herneck first described Einstein's thought experiment from Aarau of him chasing a light beam. Then Herneck referred to his earlier suggestion according to which Einstein might have thought of the speed of light already in Munich when he was twelve years old. He could find this in Bernstein's books, since Bernstein raised the question right in the introduction and continued to discuss it afterwards.[23]

Einstein's own comments on the Aarau thought experiment appear in a few sources:

**1) 1955, Autobiographische Skizze**: The thought experiment was recounted by Einstein in his *Autobiographische Skizze*, written a month before he died in March 1955:[24] "During this year in Aarau the following question came to me: if one chases a light wave with the speed of light, one would have in front of him a time independent wave field. Such a thing seems however not to exist! This was the first childish thought experiment that was related to the special theory of relativity."

**2) 1946, Autobiographical notes**: *Autobiographical notes*, written by Einstein in 1946, and published in 1949. The *Notes* were written forty to fifty years *after* the events in question.[25]

In his *Autobiographical notes* Einstein explained that much later "reflections of this type [that a "radiation must, therefore, posses a kind of molecular structure"] made it clear to me as long as shortly after 1900 […] that neither mechanics nor electrodynamics could (except in limiting cases) claim exact validity."[26] While working simultaneously on the quantum problem and the nature of radiation, and on the electrodynamics of moving bodies, "Gradually I despaired of the possibility of discovering the true laws by means of constructive efforts based on known facts." Einstein "came to the conviction that only the discovery of a universal formal principle could lead us to assured results".[27]

---

[23] Herneck, Friedrich, *Albert Einstein: ein Leben für Wahrheit, Menschlichkeit und Frieden*, 1963, Berlin: BuchVerlag der Morgen, p. 50.
[24] Einstein, Albert, "Erinnerungen-Souvenirs", *Schweizerische Hochschulzeitung* 28 (Sonderheft) (1955), pp. 145-148, pp. 151-153; Reprinted as, "Autobiographische Skizze" in Seelig Carl, *Helle Zeit – Dunkle Zeit. In memoriam Albert Einstein*, 1956, Zürich: Branschweig: Friedr. Vieweg Sohn/Europa, pp. 9-17, p. 10.
"Wenn man einer Lichtwelle mit Lichtgeschwindgkeit nachläuft, so würde man ein zeitunabhängiges Wellenfeld vor sich haben. So etwas scheint es aber doch nicht zu geben! Dies war das erste Kindliche Gedanken-Experiment das mit der speziellen Relativitätstheorie zutun hat".
[25] Norton, 2004, p. 77.
[26] Einstein, Albert ,"Autobiographical notes" In Schilpp, Paul Arthur (ed.), *Albert Einstein: Philosopher-Scientist*, 1949, La Salle, IL: Open Court. (Einstein's personal library, the Einstein Archives) (1949, pp. 1–95), pp. 48-49. Einstein owned the following copy: Schilpp, Paul Arthur (ed.), *Einstein, Scientist-Philosopher*, 1949, Library of Living Philosophers, Northwestern University, Evanston, ILL.. See details in Item 39 011, Einstein Archives, Jerusalem.

He was looking for a general principle for the electrodynamics of moving bodies, a principle of the kind one finds in thermodynamics. "After ten years of reflection such a principle resulted from a paradox upon which I had already hit at the age of sixteen: If I pursue a beam of light with a velocity $c$ (velocity of light in a vacuum), I should observe such a beam of light as an electromagnetic field at rest though spatially oscillating. There seems to be no such thing, however, neither on the basis of experience nor according to Maxwell's equations."[28]

Einstein explained his thought experiment further: "From the very beginning it appeared to me intuitively clear that, judged from the standpoint of such observer [who pursues a light beam with a velocity $c$], everything would have to happen according to the same laws as for an observer who, relative to the earth, was at rest. For how should the first observer know, or be able to determine, that he is in a state of fast uniform motion?"[29]

After speaking about the paradox in the thought experiment, Einstein goes on to say, "One sees that in this paradox the germ of the special relativity theory is already contained. Today everyone knows, of course, that all attempts to clarify this paradox satisfactorily were condemned to failure as long as the axiom of the absolute character of time, or simultaneity, was rooted unrecognized in the unconscious." However, "To recognize clearly this axiom", says Einstein, "and its arbitrary character already implies the essentials of the solution of the problem".[30]

---

[27] Einstein, 1949, pp. 48-49.
[28] Einstein, 1949, pp. 48-51.
Einstein wrote his *Autobiographical notes* at the age of 67: "I sit in order to write, at the age of sixty-seven, something like my own obituary". Einstein, 1949, pp. 2-3. From the essay that Einstein had sent his Uncle Koch, probably in 1895, we presume that he was not yet acquainted with Maxwell's equations. A year later he conceives of a thought experiment and he was to write years later about the paradox and mention these equations. Thus the sixteen year old Einstein was not yet acquainted with Maxwell's theory. He mentioned Maxwell's equations in the thought experiment because of later knowledge after he had learned these equations.
[29] Einstein, 1949, pp. 50-51. Einstein wrote, "Intuitively clear". Judged from the point of view of Einstein's creativity, Einstein reported that his primary thinking process involved visual and tactile images. It was intuitively clear to Einstein that there should be no such thing as a frozen light wave on the basis of his "invention": his ability to think in images, thought experiments, and entertain non-verbal images. He sensed that his thought experiment created a difficulty to the ether theory, even if these thought experiments could not yet be formulated in terms of words.

**3) 1947, Phillip Frank's report**: Before 1947 Einstein seemed to have told the same story about the thought experiment of age 16 to his friend Philipp Frank. Einstein very likely told Frank about this thought experiment at the same time as he wrote his *Autobiographical Notes* in 1946. Einstein told Frank the following version which appeared in the German edition of Frank's 1949 book. Frank reports that at this time (when he was in Zurich) Einstein was satisfied not with passive strange thoughts. In him already begun the seeds of ideas about the mysteries of nature, these would grow and deepen. Soon after his entry to the Polytechnic in Zurich the question, which he had already asked in the Kantonalschule in Aarau, began to trouble him more and more:[31]

"Light is something that moves and moves very quickly. The speed of light is very high, but finite and known. Now, if someone with the same speed as that of light were to run after the light, he must have remained in line with the light. For him the light would not move. But can there be a "ray of light at rest"? That seems impossible. How can something be both moving and resting? If one desires to avoid this, the movement of light should not be defined as the movement of a body or as the propagation of waves in a medium. Otherwise one would be able to pursue after light and overtake it. But if it is not there and yet it is something real, what should we actually think of a beam of light?

---

[30] Einstein, 1949, pp. 50-51.
[31] Frank, Philip, *Albert Einstein sein Leben und seine Zeit*, 1949/1979, Braunschweig: F. Vieweg, pp. 39-40.
"Aber Einstein begnügte sich um diese Zeit schon nicht mehr mit dem passive Aufnehmen fremder Gedanken. In ihm begannen bereits die Ideen zu keimen, die ihn immer tiefer in die Rätsel der natur Hineinführen sollten. Schon bald nach seinem Eintrit in das Züricher Polytechnikum drängle sich ihm die Frage immer stärker auf, die ihm eigentlich schon auf der Kantonalschule aufgegangen war. Licht ist etwas, was sich bewegt, und sich sehr rasch bewegt. Die Geschwindigkeit des Lichtes ist zwar sehr groß, aber doch endlich und bekannt. Wenn nun jemand mit derselben Geschwindigkeit dem Licht nachlaufen würde, so müßte er mit dem Licht in gleicher Linie bleiben. Für ihn würde das Licht sich nicht bewegen. Aber kann es einen "ruhenden Lichtstrahl" geben? Das scheint unmöglich. Wie kann es aber etwas geben, was zugleich ruhend und bewegt ist? Wenn man dem ausweichen will, so darf man die Bewegung des Lichtes nicht als Bewegung eines Körpers definieren oder auch nicht als fortpflanzung von Wellen in einem Medium. Denn sonst könnte man dem Licht nachlaufen und es einholen. Wenn es aber das nicht gibt und doch irgend etwas wirklich sein soll, was soll man sich unter einem Lichtstrahl eigentlich vorstellen? Diese Frage rumorte im Geist des Studenten Einstein während seiner ganzen Zeit im Polytechnikum. Er studierte alle Werke der großen Physiker darauf hin, ob sie zu der Lösung dieses Problems von der Natur des Lichtes etwas beitragen können".

This question troubled the mind of the student Einstein throughout his time in the Polytechnic. He studied all the works of the great physicists to see whether they could contribute to a solution of this problem of the nature of this light thing".

In the forward to Frank's biography Einstein wrote, "What I believe, however, I can promise the reader, is this: he will find cleaver, interesting and plausible explanations in this book, which will be at least in part new and surprising".[32] The above explanation could be one of these "new and surprising" explanations.

Frank explained how Einstein's thought experiment creates difficulties for the ether. Frank wrote that if one wants to avoid seeing a frozen light wave then the motion of light should not be defined as the propagation of waves in a medium. Otherwise one would be able to pursue after light and overtake it. If Frank reported what Einstein had told him, then it might signify that Einstein was familiar with the principle of relativity already at the age of 16, and when he was a student at the Polytechnic he could think that the ether was problematic.[33]

Frank also wrote that, soon after his entry to the Polytechnic in Zurich the question, which he had already asked in Aarau, began to trouble him more and more. Frank also said that Einstein studied all the works of the great physicists to see whether they could contribute to a solution of this problem of the nature of this light thing. We also know that during this time Einstein corresponded with his partner Mileva Marić.

**4) 1954, Biographies relying on the *Notes***: Vallentin writes, "The question that worried the youth of sixteen was: What would happen if a man tried to imprison a ray of light? The question was naturally more complex, but as scientific formulae were beyond me, it was with these simple words that Einstein explained what he himself considered to be the starting point of his work. Once he had asked himself the question, the problem haunted him. It was always present as he continued his studies at the Polytechnic and struggled with the material difficulties in his path."[34]

On December 20, 1954 Einstein wrote on Vallentin that she, "knows me superficially and what she tells about me is essentially pure invention.[35] Vallentin visited once a while Einstein and his wife Elsa in Berlin. The above description of Einstein's thought experiment was very likely taken from Einstein's *Autobiographical Notes*; in addition, Vallentin might have picked up from Frank's biography and added some "pure invention" to what Einstein had written...

---

[32] Einstein, vorwort, Frank, 1949/1979.
[33] In this year at Aarau, between 1895 and 1896, when Einstein was sixteen, Einstein presumably was also familiar with the principle of relativity in classical mechanics because while preparing for the ETH entrance examination in 1895, he had studied the German edition of Violle's textbook. Violle, Jules, *Lehrbuch der physik*, Mechanik I, Mechanik II, deutsche Ausgabe von E. Gumlich et al., 1892-1893, Berlin: J. Springer (Einstein's personal library, the Einstein Archives). Voille actually based his treatment of dynamics on the 'principle of relative motions' together with the principle of inertia. Stachel, 1989b, in Stachel 2002, p. 196.
[34] Vallentin Antonina, *Le drame d'Albert Einstein*, 1954, Paris: Plon, p. 18.
[35] Rosenkranz, Zeev and Wolf Barbara, *Albert Einstein: The Persistent Illusion of Transience*, French edition, 2007, Jerusalem: Magnes Press, p. 245.

Seelig reported, "The seeds of his famous theory, moreover, went back to his early youth during which he pondered for a long time on the image of a man who runs after a ray of light and on the situation in which a man finds himself in a falling lift. The first image of the ray of light had led to the special relativity theory, and the second of the descending lift to the general relativity theory".[36] As to the seeds of the special relativity theory, Seelig, like Vallentin, probably also took the part of the beam rider from Einstein's *Autobiographical Notes*. As to the second lift experiments Seelig could read Infeld and Einstein's book *Physik als Abenteuer der Erkenntnis*, published in 1938.[37]

**5) 1916/1945, Max Wertheimer's report**: In 1916 Wertheimer met with Einstein in order to probe the psychology of his work. He interviewed Einstein at that time, ten years after he had devised his thought experiment. Later, the interviews were published posthumously in 1945 after Wertheimer's death in 1943. Wertheimer's report seems important in that it provides an apparently independent account based on interviews much closer in time to the events.[38] Wertheimer reported:[39]

"*Act I. The beginning of the Problem*

The problem started when Einstein was sixteen years old, a pupil in the Gymnasium (Aarau, Kantonschule). […] The process started in a way that was not very clear, and is therefore difficult to describe – in a certain state of being puzzled. First came such questions as: What if one were to run after a ray of light? What if one were riding on the beam? If one were to run after a ray of light as it travels, would its velocity thereby be decreased? If one were to run fast enough, would it no longer move at all?... To young Einstein this seemed strange.

The same light ray, for another man, would have another velocity."

Wertheimer then asked Einstein, "Whether, during this period, he already had some idea of the constancy of light velocity, independent of the movement of the reference system. Einstein answered decidedly: "No, it was just curiosity. That the velocity of light could differ depending upon the movement of the observer was somehow characterized by doubt. Later developments increased that doubt' ".[40]

---

[36] Seelig, 1956, p. 70; Seelig, 1954, p. 84.
[37] Einstein, Albert und Leopold, Infeld, *Physik als Abenteuer der Erkenntnis*, 1938, Leiden: A.W. Sijthoff. (Einstein's personal library, the Einstein Archives); Einstein, Albert, and Infeld, Leopold, *The Evolution of Physics*, edited by Dr. C. P. Snow, 1938, The Cambridge library of Modern Science, Cambridge University Press, London. (Einstein's personal library, the Einstein Archives); Einstein, Albert und Infeld Leopold, *Die Evolution der Physik :von Newton bis zur Quantentheorie*, 1938/1956, Hamburg : Rowohlt. (Einstein's personal library, the Einstein Archives).
[38] Norton, 2004, p. 77.
[39] Wertheimer, Max, *Productive Thinking*, 1916/1945, New-York: Harper & Brothers, p.169.

As opposed to Einstein in his *Autobiographical Notes*, written some 30 years after Wertheimer's interview with him, the latter does not mention Maxwell's equations, nor does he mention "the absolute character of time, or simultaneity" in relation to the thought experiment. Norton writes that Wertheimer's and Einstein's 1955 reports "portray the thought experiment in its original form as much less of the well reasoned and polished display pieces that characterize Einstein's scientific writing".[41]

It appears that Phillip Frank's report concerns Einstein's pondering the problem of the optics and electrodynamics of moving bodies and thought experiment while he was a student in the Polytechnic. Einstein's own account in his *Autobiographical Notes* mixes the 16 year old thought experiment with his later understanding. However, Einstein did not give geographical "times" of "places" or temporal locations in his *Notes*; thus his *Notes* are not *a historical account* of his route to his theories, but rather a scientific explanation of his route. This explains the apparent mixture between earlier events and later understanding. Wertheimer's report is a report by someone who was not a physicist (Wertheimer was a pioneer Gestalt psychologist). Einstein explained to Wertheimer the thought experiment in a popular manner, and Wertheimer reported Einstein's intuitive explanation. Finally, Einstein's 1955 explanation from the *Skizze*, unlike the *Notes*, contains locations, names and events. The thought experiment is described within the framework of the years Einstein spent in the Kantonsschule in Aarau. He opens the *Skizze* with an account of his happy years at this Schule: it was during his time in Aarau that he imagined the childish thought experiment of chasing a light beam.

It seems that – *from the historical point of view* – Frank supplies the most important information. At first in 1895 Einstein arrived in a non-verbal manner at the thought experiment of chasing after a beam of light. This thought experiment then troubled Einstein later during his student days at the Zürich Polytechnic, and might have lead him to decide that the ether was problematic.

## 3. Magnet and Conductor Thought Experiment

In later recollections, Einstein stressed the importance of several thought experiments in his thinking that led up to the final special theory of relativity. They include his chasing a light beam thought experiment and his magnet and conductor thought experiment. They do not include any thought experiments on clocks and their synchronization.[42]

---

[40] Wertheimer, 1916/1945, p.170.
[41] Norton, 2004, p. 78.
[42] Norton, 2008, pp. 1-2.

Einstein explained in 1920 that the magnet and conductor experiment (magneto-electric induction) played a leading role for him when he established the special theory of relativity. Induction seemed to have occupied Einstein from the technical point of view and from the theoretical point of view. Einstein was fascinated with inventions and was attracted to the technical aspects of induction.

**3.1 Maxwell Equations and Induction**

In the 1920 manuscript, "Fundamental Ideas and methods of the Theory of Relativity, Presented in Their Development", Einstein spoke about the crucial role of the magnet and conductor experiment in his thought,

"When I established the theory of special relativity, the following idea on Faraday's magneto-electric induction – so far not mentioned [in the paper] – played a guiding role for me.

According to Faraday, during the relative motion of a magnet and an electric circuit, an electric current is induced in the latter. Whether the magnet is moved or the conductor doesn't matter; it only depends on the relative motion. But according to the Maxwell-Lorentz theory, the theoretical interpretation of the phenomenon is very different for the two cases:[43]

[This explanation is quite similar to the one that appears in the 1905 relativity paper…[44]]

---

[43] James Clerk Maxwell explained in an assymetric manner the magnet and conductor experiment using two types of "electromotive forces" – one for the case when the magnet is moved in the presence of a conductor, and the other for the case when the conductor is moved in a field of a magnetic force: "(7) When an electric current or a magnet is moved in presence of a conductor, the velocity of rotation of the vortices in any part of the field is altered by that motion. The force by which the proper amount of rotation is transmitted to each vortex, constitutes in this case also an electromotive force, and, if permitted, will produce currents. (8) When a conductor is moved in a field of magnetic force, the vortices in it and its neigbourhood are moved out of their places, and are changed in form. The force arising from these charges constitutes the electromotive force on a moving conductor, and is found by calculation to correspond with that determined by experiment". Maxwell, James Clerk, "On Physical Lines of Force. Part II – The Theory of Molecular Vortices Applied to Electric Currents", *Philosophical Magazine*, pp. 838-848; p. 838.
[44] If the conductor is at rest in the ether and the magnet is moved with a given velocity, a certain electric current is induced in the conductor. If the magnet is at rest, and the conductor moves with the same relative velocity, a current of the same magnitude and direction is in the conductor. However, the ether theory gives a different explanation for the origin of this current in the two cases. In the first case, according to Faraday's induction law, an electric field is supposed to be created in the ether by the motion of the magnet relative to it. In the second case, no such electric

The phenomena of magneto-electric induction forced me ["zwang mich"] to postulate the principle of (special) relativity.

[footnote] The difficulty to be overcome then lay in the constancy of the speed of light in vacuum, which I first thought would have to be given up. Only after years of groping did I realize that the difficulty rested in the arbitrariness of the fundamental concepts of kinematics"[45]

Analyzing this experiment within the framework of Maxwell's theory – as then interpreted – seemed to preclude that the mechanical (physical) principle of relativity should hold for electromagnetic phenomena. According to this principle, the laws of mechanics take the same form in all inertial reference frames. In what way then does Maxwell's theory differ from a Galilei-invariant theory (a theory which is in accord with the principle of relativity)? The answer is through the presence of the Faraday induction term, which destroys their exact Galilei invariance – thereby creating their Lorentz invariance. Hence the law of induction – or the magnet and conductor thought experiment – sets up the conflict between electrodynamics and the Galilean principle of relativity.[46]

Max Born recalled in 1955 that, "in spring 1915" he was "called to Berlin by *Planck*, to assist him in teaching. […] I was near *Planck* and *Einstein*. It was the only period when I saw *Einstein* very frequently, at times almost daily, and when I could watch the working of his mind and learn his ideas on physics and on many other subjects. […] When speaking of the physical facts which *Einstein* used in 1905 for his special relativity I said that it was the law of electromagnetic induction which seemed to have guided *Einstein* more than even *Michelson*'s experiment. Now the induction law was at that time about 70 years old – *Faraday* discovered it in 1834 – everybody had known all along that the effect depended only on relative motion, but nobody had taken offence at the theory not accounting for this circumstance".[47]

---

field is supposed to be present since the magnet is at rest in the ether, but the current results from the motion of the conductor through the static magnetic field.

[45] Einstein, Albert, "Grundgedanken und Methoden der Relativitätstheorie in ihrer Entwicklung dargestellt", 1920, Unpublished draft of a paper for *Nature* magazine, *The Collected Papers of Albert Einstein. Vol. 7: The Berlin Years: Writings, 1918–1921* (*CPAE*, Vol 7), Janssen, Michel, Schulmann, Robert, Illy, Jószef, Lehner, Christoph, Buchwald, Diana Kormos (eds.), Princeton: Princeton University Press, 1998, Doc. 31, p. 20.

[46] Jammer, Max and Stachel, John, "If Maxwell had Worked between Ampère and Faraday: An historical Fable with a Pedagogical Moral", *American Journal of Physics* 48, 1980, pp. 5-7; pp. 5-6.

[47] Born Max, *Physics in My Generation: A selection of Papers*, 1969, London: Pergamon Press, pp. 107-108; Born, Max, *Physik Im Wandel Meiner Zeit*, 1957, Germany: F. Vieweg, pp. 193-194.

Indeed the magnet and conductor thought experiment opens Einstein's 1905 relativity paper and not the famous Michelson-Morley second order in v/c ether drift experiment. In fact this latter experiment is not even mentioned in the relativity paper.

**3.2 What Prompted Einstein to Invent the Magnet and Conductor Thought Experiment?**

It appears that Einstein's thoughts about the magnet and conductor thought experiment can be traced to a number of sources.

**1) Einstein's Childhood and Technical Background**: The electro-technological environment in which Einstein grew up in Munich can explain his affinity to technical inventions. Hermann and his brother Jakob ran an electrotechnical company, "Elektrotechnische Fabrik" that produced, "Fabrikation von Dynamo-Maschinen".[48] Between 1880 and 1894, his father Herman and his uncle Jakob produced in their Munich factory electrical machinery, especially dynamos, and electrical apparatus. Jakob was granted six patents during the Munich years of the factory, and according to Pyenson all this must have established an environment and framed Einstein's mind in his development.[49]

Einstein's attention could have been drawn to the problem of unipolar induction by this environment.[50] Unipolar induction is circular motion and it is much more complicated than linear motion which Einstein finally chose.

---

See also: Born, Max, Mein Leben: *Die Erinnerungen des Nobelpreisträgers*, 1975, München: Nymphenburger Verlagshandlung GmbH; *My Life Recollections of a Nobel Laureate*, 1978, New York, Charles Scribner's Sons.
[48] Galison, 2003, p. 250.
[49] Pyenson, Lewis, *The Young Einstein: The Advent of Relativity*, 1985, Boston: Adam Hilger, p. 53.
[50] During Einstein's patent years, from June 1902 until October 1909, machines surrounded him. Indeed a letter from the Swiss Patent Office on the AEG Alternating Current Machine shows that Einstein examined the specifications of induction machines. Swiss Patent Office Letter on the AEG Alternating Current machine, December 11, 1907, *The Collected Papers of Albert Einstein. Vol. 5: The Swiss Years: Correspondence, 1902–1914* (*CPAE*, Vol. 5), Klein, Martin J., Kox, A.J., and Schulmann, Robert (eds.), Princeton: Princeton University Press, 1993, Doc. 67. Later we recognize Einstein's tendency towards technology and especially induction. Between 1907 and 1910 Einstein and the Habicht brothers (Conrad and Paul) were experimenting on induction machine ("The Maschinchen" – "little machine") for measuring small voltages by multiplication (a detector). "Einstein's 'Maschinchen' for the Measurement of Small Quantities of Electricity", *CPAE*, Vol. 5, pp. 51-55. Einstein's part in experimenting with the little machine had probably its roots in Einstein's youth environment. The Habicht brothers and Einstein's attempts to perfect the induction machine lasted several years. Seelig says that nothing world-shattering ever came from it, but as "team-work it helped to tighten the bond friendship. Einstein

Einstein also studied *Faraday's induction* at secondary school, the Luitpold Gymnasium. Einstein studied from a physics book by Dr. Jos Krist, *Anfangsründe der Naturlehre für die Unterclassen der Realschulen*. Krist in this book explained "Magneto-induction", and then associated it with Faraday's induction.[51]

**2) The Polytechnic and August Föppl's book on Maxwell's theory**: Heinrich Friedrich Weber's courses at the Polytechnic could also influence Einstein. Einstein took these courses and laboratories with Weber: "Electrotechnics – electrical oscillators – electro-technical laboratory – scientific work in the physics laboratory – introduction to electromechanics – alternating currents – the system of absolute electrical calibration".[52] Thus the combination of Einstein's work in Weber's laboratories and his technical background from home could lead him to invent the Magnet and Conductor thought experiment.

Reiser writes, "Just then his one great love among the sciences was physics. But the scientific courses offered to him in Zürich soon seemed insufficient and inadequate, so that he habitually cut his classes. His development as a scientist did not suffer thereby. With a veritable mania for reading, day and night, he went through the works of the great physicists – Kirchhoff, Hertz, Helmholtz, Föppl".[53] It is important to stress that Einstein never mentioned August Föppl among his reading list.

---

had lodged a patent for this jointly developed apparatus and compiled the particulars. Seelig, 1956, p. 60; Seelig, 1954, p. 73.
Later, in the 1908 issue of the *Physikalische Zeitschrift* appeared Einstein's own detailed description of the means by which with this induction machine could be used to measure the minutest electric voltages down to the range of some 0.0005 volts. Two years later the Habicht brothers described the "Einstein-Habicht Potential-Multiplikator", which after many attempts they had built, and had even carried out the experiments together with Einstein in Zürich University laboratory. Finally they applied for a patent in order to produce this new measurement apparatus in their own factory. Seelig, 1956, pp. 60-61; Seelig, 1954, p. 73.
[51] Krist, Jos. Dr., *Anfangsründe der Naturlehre für die Unterclassen der Realschulen*, 1891, Wien: Wilhelm Braumüller, K. U. K Hof. Und Universitäts-Buchhändler, p. 94. "*In einer geschlossenen Drahtrolle entsteht ein Inductionsstrom, wenn 1. ihr ein ein Magnetpol genähert, 2. in ihr ein Magnetpol hervorgerufen oder nur verstärkt wird. Ebenso entsteht ein Strom, jedoch von entgegengesetzter Richtung, wenn 3. der Magnetpol entfernt wird oder 4. seinen Magnetismus ganz oder iheilweise verliert. Diese Induction heißt M a g n e t o-I n d u c t i o n (F a r a d a y 1831)*".
[52] Seelig, 1956, p. 25; Seelig, 1954, p. 30.
[53] The journalist Rodulf Kayser, Einstein's son-in-law, the husband of Einstein's stepdaughter Ilse (writing under the pseudonym Anton Reiser), wrote a biography with Einstein's approval and his so-called cooperation. However, Einstein did not consider Reiser's biography as a most reliable book. Einstein wrote in the preface to Reiser's book: "The author of this book is one who knows me rather intimately in my endeavor, thoughts, beliefs – in bedroom slippers […] I found the facts of this book duly accurate, and its characterization, throughout, as good as might be expected of one who is perforce himself, and who can no more be another than I can". Reiser,

Nevertheless, Holton, very likely relying on Reiser's report, refers to another source for the magnet and conductor thought experiment: the young Einstein read during his student days at the Polytechnic, Föppl's 1894 *Introduction to Maxwell's Theory of Electricity*.[54]

The fifth main section of Föppl's book was entitled *The Electrodynamics of Moving Conductors*, *Die Elektrodynamik bewegter Leiter*. In this section, on pages 307-356, there was a chapter entitled, "Electromagnetic-Force Induction by Movement". The first paragraph in this chapter is "Relative and Absolute Motion in Space". It starts in the following way,

"The discussion of kinematics, namely of the general theory of motion, usually rest on the axiom that, in the relationship of bodies to one another only relative motion is of importance. There can be no recourse to an absolute motion in space since there is absent any means to find such a motion if there is no reference object at hand from which the motion can be observed and measured".[55]

Föppl continues a few lines later in this way: "When in the following we make use of laws of kinematics for relative motion, we must proceed with caution. We must not consider it as a priori settled that it is, for example, all the same whether a magnet [moves] in the vicinity of a resting electric circuit or whether it is the latter that moves while the magnet is it rest." Holton writes, "This, we recall, describes precisely the experimental situation with which Einstein's paper starts".[56] According to Holton we have good reasons to believe that Einstein had read August Föppl's book, which contains a version of unipolar induction: the generation of a current on the conductor for the case in which the conductor and the magnet are in relative rotation motion. However, we do not know when Einstein read Föppl's book, since he nowhere mentions it in his writings; and thus it is difficult to say whether this book inspired him to invent the magnet and conductor thought experiment.

**3) A Conversation with Besso:** In a letter of August 3, 1952 Besso recounted,[57]

---

Anton [Rudolf Kayser] *Albert Einstein: A Biographical Portrait*, 1930/1952, New York: Dover. (Einstein's personal library, the Einstein Archives), p. 49; Reiser, Anton [Rudolf Kayser], *Albert Einstein, Ein Biographisches Porträt*, 1930/1997, New-York: Albert & Carles Boni (A German manuscript which was never published because Einstein refused to give permission for publishing the book in German).
[54] Föppl, August, *Einführung in die Maxwell'sche Theorie der Elektricität*, 1894, Leipzig: Teubner.
[55] Holton, 1973/1988, p. 221.
[56] Holton, 1973/1988, p. 221.
[57] Besso to Einstein, August 3, 1952, in Einstein and Besso, 1971, Speziali, *Correspondance*, Letter 188, pp. 477-478; p. 478.
"Ein anderes Märlein: aus meiner Meinung doch an der SRT beteiligt gewesen zu sein: indem mir, als Elektrotechniker, nahe liegen musste, das, was im Rahmen der Maxwellschen Theorie, je nachdem der Induktor eines Alternators ruht oder rotiert,

"Another little story of mine concerning my view that I had participated in [the formulation of] the special theory of relativity. It seemed to me, as an electrical engineer, I must have brought up, in conversations with you, the question, within the context of Maxwell's theory, of what is induced in the inductor of an alternator; depending on whether it is at rest or rotating, there is induced in the inductive part an electromotive [i.e., a magnetic] force or a [purely] electric one, as a peculiarly practical anticipation of the relativistic point of view […] That this somehow still resonates emotionally within me is demonstrated by the confusing sentence structure. May Spinoza and Freud watch over me".

Besso first studied mathematics and physics at Trieste, and then at the University of Rome (1891-1895) there he took courses in mathematics and physics, and thus he could have learned Maxwell's theory there, but there is no evidence for this. On the advice of his uncle David, who taught mathematics at the University of Modana (Italy), he left for Zurich and enrolled in October 1891 in the mechanics section of the Federal Polytechnic school, the ETH. After four years of study he obtained his diploma in mechanical engineering, and soon afterwards, a position in an electrical-machinery factory in Zurich. However, he could *not* have been acquainted with Maxwell's theory from his studies in the Polytechnic.[58] Weber did not teach Maxwell's theory in the Polytechnic. Dr. Joseph Sauter, Weber's assistant and later Einstein's colleague at the Bern Patent Office, recalled that "this theory [Maxwell's] was not yet on the official program of the Zürich Polytechnic School".[59]

Therefore, it was probably Einstein's self-reading about Maxwell's theory, who explained to Besso about this theory. Only after such explanation could Besso "im Rahmen der Maxwellschen Theorie" (within the context of Maxwell's theory) refer to his technical work and speak with Einstein about induction (of which Einstein had already read about either in Krist's book or in Föppl's book).

## 4. Ether drift and Michelson and Morley's experiment

### 4.1 Einstein Designs Ether Drift Experiments between 1899 and 1901

A year after inventing the chasing a light beam thought experiment, Einstein was a student at the Zurich polytechnic. By 1899, he was interested in ether drift

---

im induzierten Teil als elektromotorische oder als elektrische Kraft auftritt, als eine eigentümliche praktische Vorwegnahme der Relativitätsauffassung – in das Gespräch hineingebracht zu haben... Dass in mir irgendwie noch ein Affekt mitschwingt, zeigt sich am verworrenen Satzbau. Spinoza und Freud halten mich mir gegenüber wach".
[58] Speziali, Pierre, "Einstein writes to his best friend", in French, A. P. (ed), *Einstein A Centenary Volume*, 1979, London: Heinemann for the International Commission on Physics Education, pp. 263-269; pp. 263-264.
[59] Sauter, Joseph, "Comment j'ai appris à connaître Einstein", 1960, in Flückiger, Max (1960/1974), *Albert Einstein in Bern*, 1974, (Switzerland: Verlag Paul Haupt Bern), p. 154.

experiments, and appears to have designed one. According to the 1922 Kyoto lecture Einstein said: [60]

"Then I myself wanted to verify the flow of the ether with respect to the Earth, in other words, the motion of the Earth. When I first thought about this problem, I did not doubt the existence of the ether or the motion of the Earth through it. I thought of the following experiment using two thermocouples: Set up mirrors so that the light from a single source is to be reflected in two different directions, one parallel to the motion of the Earth and the other antiparallel. If we assume that there is an energy difference between the two reflected beams, we can measure the difference in the generated heat using two thermocouples".

The above paragraph may well have described the idea that Einstein had in 1899 during a visit in Aarau. He wrote Marić on September 10, 1899, "In Aarau I had a good idea for investigating the way in which a body's relative motion with respect to the luminiferous ether affects the velocity of the propagation of light in transparent bodies. I even came up with a theory about it that seems quite plausible to me."[61]

Einstein wrote Marić probably on September 28, 1899, "I also wrote to Professor Wien in Aachen about my paper on the relative motion of the luminiferous ether relative to ponderable matter, the paper which the 'boss' handled in such an offhanded fashion. I read a very interesting paper by Wien from 1898 on this subject."[62]

Reiser reported,[63]

"He encountered at once, in his second year of college, the problem of light, ether and the earth's movement. This problem never left him. He wanted to construct an apparatus which would accurately measure the earth's movement against the ether. That his intention was that of other important theorists, Einstein did not yet know. He was at that time unacquainted with the positive contributions, of some years back, of the great Dutch physicist Hendrick Lorentz, and with the subsequently famous attempt of Michelson. He wanted to proceed quite empirically, to suit his scientifically feeling of the time, and believed that an apparatus such as he sought would lead him to the solution of a problem, whose far-reaching perspectives he already sensed.

---

[60] Einstein, Albert (1922), "How I Created the Theory of Relativity, translation to English by. Yoshimasha A. Ono, *Physics Today* 35, 1982, pp. 45-47 (notes taken originally in Japanese by Jun Ishiwara who sat in the audience and heard Einstein talk, and not approved by Einstein) p. 46. Jun Ishiwara was a physicist who visited Europe and met many of the physicists involved in the development of modern physics, and the person who invited Einstein to visit Kyoto in Japan and arranged the details of the trip.
[61] Einstein to Marić, September 10, 1899, *CPAE*, vol 1, Doc. 54; Renn, Jürgen and Schulmann, Robert, *Albert Einstein Mileva Marić The love letters*, translated by Shawn Smith, 1992, Princeton: Princeton University Press, letter 10.
Stachel thinks that the reference to the sentence, "the velocity of the propagation of light in transparent bodies" suggests that Einstein may have had in mind some variant of Fizeau's 1851 water-tube experiment. Stachel, 1987, in Stachel, 2002, p. 173.
[62] Einstein to Marić, Sptember 28?, 1899, *CPAE*, vol 1, Doc. 57; Renn and Schulmann, 1992, letter 11.
[63] Reiser, 1930, p. 52.

But there was no chance to build this apparatus. The skepticism of his teachers was too great, the spirit of enterprise too small. Albert has thus to turn aside from his plan, but not to give it up forever".

According to Reiser above, while a student at the Polytechnic, Einstein "wanted to construct an apparatus to accurately measure the earth's movement against the ether". However, he was unable to do so because the skepticism of his teachers was too great. Thus the "boss" referred to in the letter to Marić is very likely Weber, Einstein's supervisor in physics.[64]

Wien's 1898 paper was his report to the Society of German Scientists and Physicians, "On questions concerning the translatory motion of the luminiferous ether". Wien discussed both Hertz's concept of moving ether and Lorentz's concept of immobile ether. He also briefly considered 13 experiments bearing on the question, the last of which was the Michelson-Morley experiment. It is reasonable to conjecture that Einstein read this account in 1899, and knew something about the experiment from this paper.[65]

Two years later, after graduating the Polytechnic, Einstein appeared to have designed a second ether drift experiment. On December 19, 1901 he wrote Marić,[66]

"I spent all afternoon with Professor Alfred Kleiner in Zurich telling him my ideas about the electrodynamics of moving bodies, and we talked about all sorts of other physics problems. He's not quite as stupid as I'd thought' and moreover, he's a good fellow. […] He advised me to publish my ideas on the electromagnetic theory of light of moving bodies along with the experimental method. He found the method I have proposed to be the simplest and most appropriate one imaginable. I was quite happy about the success. I will write the paper in the next few weeks for sure".

But Einstein was still not ready to publish his ideas on electrodynamics. The experimental method is presumably the method for investigating the motion of matter with respect to the ether mentioned in Einstein's letter to Marcel Grossmann of September 1901.[67] Einstein wrote that on "the investigation of the relative motion of matter with respect to the luminiferous ether, a considerably simpler method had occurred to me, which is based on customary interference experiments". Einstein promised Grossman that when they saw each other, "I will tell you about it".[68]

---

[64] Renn and Schulmann, 1992, notes 3, p. 85.
[65] Wien, Wilhelm, "Über die Fragen, welche die translatorische Bewegung des Lichtäthers betreffen", *Annalen der Physik und Chemie* 65 (1898), pp. i-xvii.
[66] Einstein to Marić, letter 47, Dec 19 1901, *CPAE*, Vol 1, Doc. 130, p. 328, Renn and Schulmann, 1992, letter 47.
[67] Renn and Schulmann, 1992, note 4, p. 100.

Wertheimer described Einstein's efforts at finding experimental methods for detecting the motion of the earth with respect to the ether:[69]

"What *is* 'the velocity of light'? If I have it in relation to something, this value does not hold in relation to something else which is itself in motion. (Puzzling to think that under certain conditions light should go more quickly in one direction than another.) If this is correct, then consequences would also have to be drawn with reference to the earth, which is moving. There would then be a way of finding out by experiments with light whether one is in a moving system! Einstein's interest was captured by this; he tried to find methods by which it would be possible to establish or to measure the movement of the earth – and he learned only later that physicists had already made such experiments".

**4.2 The Role of the Michelson-Morley Experiment on the Discovery of Relativity**

In 1901 Einstein wrote Marić, "I want to get down to business now and read what Lorentz and Drude have written about the electrodynamics of moving bodies".[70] Therefore in 1902 Einstein already read Lorentz's 1895 *Versuch* and from it he probably learnt about the Michelson-Morley experiment (If he had not done so earlier from Wien).

Michelson and Morley's experiment became more and more famous at that time. The Michelson experiment[71] showed that one could not disclose earth's motion relative to the ether up to the second order. And the Michelson and Morley experiment demonstrated even more precisely that earth's motion through the ether did not produce any perceptible effect on the measuring instruments.

Michelson's experimental result is usually *cited* as preliminary problem that demanded the abandoning of the ether and, which finally have *led* to Einstein's solution. According to this scenario, which usually appears in textbooks, it appears obvious that Einstein must have based himself on Michelson's experimental result on his pathway to special relativity; and if he did not explicitly use this experimental result, at least he knew for sure about the famous experiment prior to publication of the relativity paper. In 1949 Robert A. Millikan recollected memories about his mentor Michelson and told this story.

---

[68] Einstein to Grossmann, probably on September 6, 1901, *CPAE*, vol 1, Doc. 122.
[69] Wertheimer, 1916/1945, p. 169.
[70] Einstein to Marić, letter 131, Dec 28 1901, *CPAE*, Vol 1, Doc. 130, p. 328, Renn and Schulmann, 1992, letter 48.
[71] Michelson, Albert Abraham, "The Relative Motion of the Earth and the Luminiferous Ether", *American Journal of Science* 22, 1881, pp.120-129; Michelson, Albert Abraham, and Morley Edward W., "On the Relative Motion of the Earth and the Luminiferous Ether", *American Journal of Science* 34, 1887, pp. 333-345.

In the 1949 Einstein issue of *Reviews of Modern Physics*, celebrating Einstein's seventieth birthday, Millikan wrote in his paper, "Albert Einstein on his Seventieth Birthday" the following, "In the case of relativity the prime experimental builder had been my own chief at the University of Chicago, Albert A. Michelson, who made his first experiment on aether-drift at Berlin in 1881, only two years after he had risen to fame by making in 1879 a very great improvement upon Foucault rotating mirror method of determining the speed of light. With aid from Alexander Graham Bell he spent the next two years in Europe and in Paris set up the first 'Michelson interferometer'. The next year he made with it the earliest attempt at an aether-drift determination. In his brief report on this experiment in the American Journal of Science 22, 120, (1881), he is so sure of the correctness of the negative result obtained that he asserts that, in spite of the crudity of his apparatus, 'the hypothesis of a stationary aether is thus shown to be incorrect'.

But it was not until 1887 that this experiment, repeated at Case School of Applied Science with great care and refinement by Michelson and Morley, began to take its place as the most famous and in many ways the most fundamentally significant experiment since the discovery of electromagnetic induction by Faraday in 1831. The special theory of relativity may be looked upon as starting essentially in a generalization from Michelson's experiment".[72]

Millikan then went on to say, "But this experiment, after it had been performed with such extraordinary skill and refinement by Michelson and Morley, yielded with great definiteness the answer that there is no such time-difference and therefore no observable velocity of the earth with respect to the aether. That unreasonable, apparently inexplicable experimental fact was very bothersome to 19$^{th}$ century physics and so for almost twenty years after this fact came to light physicists wandered in the wilderness in the disheartening effort to make it seem reasonable. Then Einstein called out to us all, 'Let us merely accept this as an established experimental fact and from there proceed to work out its inevitable consequences', and he went at that task himself with an energy and capacity which very few people on earth possess. Thus was born the special theory of relativity." [73]

Holton cites this paragraph of Millikan in his book *Thematic Origins of Scientific Thought*, and asks: "How important were experiments to Einstein's formulation of his 1905 paper on relativity? What role did the Michelson experiments play?" Holton says that "Einstein himself made different statements about the influence of the Michelson experiments, ranging from 'there is no doubt that Michelson's experiment was of considerable influence on my work…' to 'the Michelson-Morley experiment had a negligible effect on the discovery of relativity'."[74]

---

[72] Millikan, Robert A., "Albert Einstein on His Seventieth Birthday", *Reviews of Modern Physics* 21, 1949, pp. 343-345; p. 343.
[73] Millikan, 1949, pp. 343-344.

There are good reasons to ask: why is Michelson and Morley's experiment so intimately connected with the special theory of relativity? Holton answered this question as follows,[75]

"For years after Einstein's first publication no new experimental results came forth which could be used to 'verify' his theory in the way most physicists were and still are used to look for verification. As Max Planck noted in 1907, Michelson's was then still regarded as the only experimental support. [...]

In retrospect it seems therefore inevitable that during the decade following Einstein's 1905 paper there occurred – especially in the didactic literature – a symbiotic joining of the puzzling Michelson experiment and the all-but-incredible relativity theory. The undoubted result of Michelson's experiments could be thought to provide an experimental basis for the understanding of relativity theory, which otherwise seemed contrary to common sense itself; the relativity theory in turn could provide an explanation of Michelson's experimental result in a manner not 'artificial' or 'ad hoc' as reliance on the supposed Lorentz- FitzGerald contraction was a widely felt to be. It proved to be a long lasting marriage".

**4.3 Einstein and Michelson in America 1930**

Albert Einstein first met Michelson on his trip to America in 1930. Einstein first came to California on December 31, 1930. He arrived in San Diego on December 31, 1930 on the steamship *Belgenland*, together with his wife Elsa, mathematician Walter Mayer, and Secretary Helen Dukas. They were driven in a motorcade two hundred miles north to Pasadena, where Einstein spent the next three months at Caltech's Norman Bridges Laboratory in association with some of the leading figures in American science.[76]

In 1921 Millikan had assumed the role of director of the Norman Bridges Laboratory of Physics as well as chairman of the Executive Council of the former Throop Polytechnic Institute, which had been renamed California Institute of Technology in 1920. Soon after his arrival Millikan announced his desire to put physics on the map in southern California, and initiated a visiting-scholars program. Distinguished American physicist Albert A. Michelson of the University of Chicago, the first American Jew to win the Nobel Prize, was immediately recruited. Einstein's visits to

---

[74] Holton, 1973/1988, pp. 280-281.
[75] Holton, 1973/1988, p. 287.
[76] Kramer, William, M., *A Lone Traveler, Einstein in California*, 2004, California: Skirball, pp. 2, 4.

Caltech capped Millikan's campaign to make the university one of physics capitals of the world.[77]

On January 15, 1931 distinguished scholars, researchers and other guests assembled to honor Einstein at the Athenaeum, the elegant dining hall of the Institute's faculty.[78] Among them was Michelson, 79 years old, and weaken by a serious stroke. The picture taken on the occasion of the meeting shows the frail old Michelson standing next to Einstein. This was Michelson's last public appearance; he died three months later.[79]

Holton wrote that "Millikan set the stage with some remarks on what he saw to be the characteristic features of modern scientific thought. It is, in fact, largely, the very same material Millikan was to republish 18 years later in 1949 as part of his introduction for the Einstein issue of the *Review of Modern Physics*. But after the sentence, 'Thus was born the special theory of relativity ', Millikan went on to say in 1931 'I now wish to introduce the man who laid its experimental foundations, Professor Albert A. Michelson…".[80]

Millikan had paid tribute to Michelson as the man who laid the experimental foundations to special relativity, and Michelson was now going to thank him for this great honor. Einstein cooperated in this tribute without questioning this narrative.

Michelson thus responded, "I consider it particularly fortunate for myself to be able to express to Dr. Einstein my appreciation of the honor and distinction he has conferred upon me for the result which he so generously attributes to the experiments made half a century ago in connection with Professor Morley, and which he is so generous as to acknowledge as being a contribution on the experimental side which led to his famous theory of relativity".[81]

Holton wrote that in fact Michelson was known to be no friend of relativity, the destroyer of the ether. Like so many others, Michelson was convinced that his own ill-fated experiments were the basis for the theory. Einstein reminisced later that Michelson "told me more than once that he did not like the theories that had followed from his work", and that he had told Einstein he was a little sorry that his own work had started this "monster".[82]

---

[77] Kramer, 2004, p. 10.
[78] Kramer, 2004, p. 11.
[79] Holton, Gerald, "Einstein and the 'Crucial' Experiment", *American Journal of Physics* 37, 10, 1969, pp. 968-982; p. 973.
[80] Holton, 1969, p. 973.
[81] Holton, 1969, p. 973.
[82] Holton, 1969, pp. 972-973.

Einstein told Charles Nordmann during his visit to Paris in 1922 that Michelson once said: "If I had guessed that the results of my experiment would lead to this thing, I really believe that I never would have done it".[83]

Back to the grand dinner honoring Einstein in Pasadena; Einstein now had to respond to Michelson's warm words. What happened now, says Holton, is widely known from the account given in Michelson's biography by B Jaffe, the only one then available when Holton wrote his paper in 1969,[84]

Einstein made a little speech, Seated near him were Michelson, Millikan, Hale and other eminent men of science.[85] Albert Einstein was introduced as "the greatest physicist of this period, the one who has inspired and guided all the major lines of physical investigation during the past 25 years". As the guest of honor walked to the podium, the audience rose to its feet and applauded in a prolonged ovation. Dressed in a black swallowtail coat with a large white gardenia pinned to his left lapel Einstein started to speak, "My dear friends, from far away I have come to you, but not to strangers. I have come among men who for many years have been true comrades with me in my labors".[86]

Einstein then turned to Michelson and said: [87]

"You my honored Dr. Michelson, began with this work when I was only a little youngster hardly three feet high. It was you who led the physicists into new paths and through marvelous experimental work paved the way for the development of the theory of relativity. You uncovered an insidious defect in the ether theory of light, as it then existed, and stimulated the ideas of H. A. Lorentz and FitzGerald, out of which the Special Theory of Relativity developed. Without your work this theory would today be scarcely more than an interesting speculation: It was your verifications which set the theory on a real basis".

Michelson was deeply moved. There could be no higher praise for any man, says Jaffe.[88] Einstein's remarks were broadcasted live on radio throughout the United States and in Germany over the Columbia Broadcasting System. [89]

---

[83] Nordmann, Charles, "Einstein Expose et Discute sa Théorie", *Revue des Deux Mondes*, 1922, Tome Neuvième, Paris, pp. 130-166; p. 142.
"A propos de l'expérience de Michelson, Einstein m'a raconté de puis, que le célébre physicien américain lui avait dit un jour: "Si j'avais pu deviner qu'on tirerait des résultats de mon experience tout ce qu'on en tiré, je crois bien que je ne l'aurais j'amais faite".
[84] Jaffe, B., *Michelson and the Speed of Light*, 1960, New York: Doubleday & co., Inc in Holton, 1969, p. 973.
[85] Holton, 1969, p. 973.
[86] Kramer, 2004, p. 12.
[87] Kramer, 2004, p. 11; Holton, 1969, p. 973.
[88] Holton, 1969, p. 973.
[89] Kramer, 2004, p. 12.

Speculation swirled about the coming together of Einstein, Michelson, and Millikan. The *New York Times* wrote on December 1930, "Michelson, far advanced in years, still seeking absolute figures for the speed of light: Millikan, delving into the question of the creation of new worlds out of radiant energy; and Einstein, pad in hand, jotting down figures and symbols".[90]

## 4.4 Einstein Different Statements as to the Role that Michelson's Experiment Played in his Development

Over the years Einstein has been asked many times whether the Michelson experiments influenced his thought. Holton differentiates "between the statements Einstein made in direct response to repeated requests to deal with the possible genetic role of the Michelson experiment, and the rather different statement he made whether he volunteered any comments concerning the genesis of relativity (in which case Einstein almost always spoke only about the experiment of Fizeau and the aberration measurements, insofar as he spoke of measurements at all)".[91]

It appears that when Einstein worked on the problems of electrodynamics of moving bodies he took for granted that one did not have to go over every ether-drift experiment in order to convince oneself about the non-existence of absolute motion. Einstein did not remember whether he knew of Michelson's experiment, because for him it was not a problem of ether-drift experiments. He was rather concerned with heuristic principles.

Wertheimer explained the situation after talking to Einstein,[92]

"He felt that the trouble went deeper than the contradiction between Michelson's actual and the expected result.

He felt that a certain region in the structure of the whole situation was in reality not as clear to him as it should be, although it had hitherto been accepted without question by everyone, including himself. His proceeding was somewhat as follows: there is a time measurement while the crucial movement is taking place. 'Do I see clearly" he asked himself, 'the relation, the inner connection between the two, between the measurement of time and that of movement? Is it clear to me how the measurement of time works in such a situation?' And for him this was not a problem with regard to the Michelson experiment only, but a problem in which more basic principles were at stake".

In a letter in honor of Michelson's 100[th] birthday and dated December 19, 1952, Einstein describes the influence of the Michelson-Morley experiment on him, "I am not sure when I first heard of the Michelson experiment or its more precise repetition by Michelson and Morley. I was not conscious that it influenced me directly during the seven and more years that the development of the Special Theory of Relativity had been my entire life; for I had taken it for granted as being true".[93]

---

[90] "Michelson Forgets his 78[th] Birthday; Scientist Awaits Einstein in California and Continues Lightspeed Quest", *New York Times*, December 30; Kramer, 2004, p. 12
[91] Holton, 1969, p. 969.
[92] Wertheimer, 1945, p. 174.

The typescript note is in English, with German corrections in Einstein's hand. A handwritten notation corrects the stricken typescript: "My thought was more indirectly influenced by the famous Michelson-Morley experiment. I learned of it through Lorentz's path breaking investigation on the electrodynamics of moving bodies (1895), of which I knew before the establishment of the special theory of relativity. […] My direct path to the sp. th. rel. was mainly determined by the conviction that the electromotive force induced in a conductor moving in a magnetic field is nothing other than an electric field. But the result of Fizeau's experiment and the phenomenon of aberration also guided me."[94]

Holton started his 1969 paper with a letter dated 2 February 1954, about a year before Einstein's death, from Francis Garvin Davenport of the Department of History of Monmouth College, Illinois. Davenport wrote to Einstein that he was looking into evidence that Michelson had "influenced your thinking and perhaps helped you to work out your theory of relativity". Holton wrote that Davenport, not being a scientist, asked Einstein for "a brief statement in nontechnical terms, indicating how Michelson helped to pave the way, if he did, for your theory". Holton reports that Einstein answered very soon after receipt, on 9 February 1954,[95]

"Before Michelson's work it was already known that within the limits of the precision of the experiments there was no influence of the state of motion of the coordinate system on the phenomena, resp. their laws. H.A Lorentz has shown that this can be understood on the basis of the formulation of Maxwell's theory for all cases where the second power of the velocity of the system could be neglected (effects of the first order).

According to the state of the theory it was, however, natural to expect that this independence would not hold for effects of second and higher orders. To have shown that such expected effect of the second order was *de facto* absent in one decisive case was Michelson's greatest merit."

Einstein next explained to Davenport that Michelson's experiment had a great effect in extending the range of validity of the principle of relativity,[96]

"This work of Michelson, equally great through the bold and clear formulation of the problem as through the ingenious way by which he reached the very great required precision of measurement, is his immortal contribution to scientific knowledge. This contribution was a new strong argument for the non-existence of 'absolute motion', resp. the principle of special relativity which, since Newton, was never doubted in Mechanics but *seemed* incompatible with electo-dynamics."

---

[93] Quated in Norton, 2004, p. 82.
[94] Norton, 2004, pp. 49-50.
[95] Holton, 1969, p. 699.
[96] Holton, 1969, p. 699.

Einstein then told Davenport that Michelson's experiment did not have an effect on his own development and pathway toward the Special Theory of Relativity,[97]

"In my own development Michelson's result has not had a considerable influence. I even do not remember if I knew of it at all when I wrote my first paper on the subject (1905). The explanation is that I was, for general reasons, firmly convinced that there does not exist absolute motion and my problem was only how this could be reconciled with our knowledge of electro-dynamics. One can therefore understand why in my personal struggle Michelson's experiment played no role or at least no decisive role."

Einstein ended his letter by giving Davenport the permission to publish the letter,[98]

"You have my permission to quote this letter. I am also willing to give you further explanations if required". Holton said that this is a thoughtfully composed reply, and the last letter of Einstein he had been able to find on this subject.

**4.5 Robert Shankland's Interviews with Einstein on Michelson's Experiments**

Between 1950 and 1954 Einstein pronounced his views regarding the Michelson experiments to Robert Shankland. "The first visit to Princeton to meet Professor Einstein was made primarily", so wrote Shankland, "to learn from him what he really felt about the Michelson-Morley experiment, and to what degree it had influenced him in his development of the Special Theory of Relativity.[99]

Shankland reported, on February 4 1950 he met Professor Einstein at his office in Princeton, "When I asked him how he had learned of the Michelson-Morley experiment, he told me that he had become aware of it through the writings of H. A. Lorentz, but *only after 1905* had it come to his attention! 'Otherwise', he said, 'I would have mentioned it in my paper'. He continued to say the experimental results which had influenced him most were the observations on stellar aberration and Fizeau's measurements on the speed of light in moving water. 'They were enough', he said."[100]

---

[97] Holton, 1969, p. 699.
[98] Holton, 1969, p. 699.
[99] Shankland, Robert, "Conversations with Albert Einstein I/II" *American Journal of Physics* 31, 1963, pp. 47-57; 41, 1973, pp. 895-901; p. 47.
See also: Shankland, Robert, "Comment on 'Conversations with Albert Einstein. II'", *American Journal of Physics* 43, 1974, p. 464 and: Shankland, Robert, "Michelson-Morley Experiment", *American Journal of Physics* 32, 1964, pp. 16-35.
[100] Shankland, 1963/73, p. 48.

In light of the importance of Fizeau's experiment and aberration in Einstein's thought he was convinced that absolute motion did not exist when one examined first order ether-drift experiments. For him this was enough.

On October 24, 1952, Shankland walked down Mercer Street to Professor Einstein's home. He went up to his study, and there "I asked Professor Einstein where he had first heard of Michelson and his experiment. Einstein replied, 'This is not so easy, I am not sure when I first heard of the Michelson experiment. I was not conscious that it had influenced me directly during the seven years that relativity had been my life. I guess I just took it for granted that it was true".[101]

After 1905 Einstein wrote expositions of special relativity and in most of them mentioned the Michelson-Morley experiment. Shankland wrote, "However, Einstein said that in the years 1905-1909, he thought a great deal about Michelson's result, in his discussions with Lorentz and others in his thinking about general relativity. He then realized (so he told me) that he had also been conscious of Michelson's result before 1905 partly through his reading of the papers of Lorentz and more because he had simply assumed this result of Michelson to be true".[102]

According to Shankland Einstein said on Michelson that he was "a great genius – he will always be thought so in this field". Shankland wrote, "Einstein added that it was very remarkable that Michelson with little mathematics or theoretical training and without the advice of theoretical colleagues could devise the Michelson-Morley experiment. Michelson's instinctive feeling for the essentials of a crucial experiment without completely understanding the related theories, Einstein considered the surest sign of his genius. This he feels was in large measure due to Michelson's artistic sense and approach to science especially his feeling for symmetry and form. Einstein smiled with pleasure as he recalled Michelson's artistic nature – here there was kindred bond. The artist was greatly in evidence in the Michelson-Morley experiment, Einstein remarked."[103]

**4.6 The Dayton Miller Experiments**

**4.6.1 Einstein is willing to give up his Theory of Relativity if Miller is Right**

The negative result of the Michelson-Morley experiment stimulated many repetitions of this experiment during the next fifty years, especially in light of the implications of Einstein's special theory of relativity. All trials of this experiment yielded a null result within the accuracy of the observations. However, one repetition of the experiment performed by Dayton Clarence Miller appeared to be perplexing. Miller observed very small fringe displacements, being on average only about 1/13 of those predicted by the ether theory for the 30 km/sec average velocity of the earth in its orbit.[104]

---

[101] Shankland, 1963/73, p. 55.
[102] Shankland, 1963/73, p. 55.
[103] Shankland, 1963/73, p. 49.

Einstein expressed his opinion about the Miller experiments on different occasions. In an interview with the BBC in 1966, Karl Popper told an anecdote about Einstein: "When D. C. Miller, who had always been an opponent of Einstein, announced that he had overwhelming experimental evidence against special relativity, Einstein at once declared that if these results should be substantiated he would give up his theory. At the time some tests, regarded by Einstein as potential refutations, had yielded favorable results, and for this and other reasons many physicists were doubtful about Miller's alleged refutations. Moreover, Miller's results were regarded as quantitatively implausible. They were, one might say, neither here nor there. Yet Einstein did not try to hedge. He made it quite clear that, if Miller's results were confirmed, he would give up special relativity, with it, general relativity also".[105]

It turns out that in different sources we can find quite the same reaction of Einstein to Miller's experimental result. Klaus Hentschel brings quite a few examples in his paper "Einstein's Attitude Towards Experiments: Testing Relativity Theory 1907 – 1927". One example: in a statement from January 19, 1926 on Miller's experiments, 'Meine Theorie und Millers Versuche', Einstein wrote:[106]

"If the results of the Miller experiments were to be confirmed, then relativity theory could not be maintained, since the experiments would then prove that, relative to the coordinate systems of the appropriate state of motion (the Earth), the velocity of light in a vacuum would depend upon the direction of motion. With this, the principle of the constancy of the velocity of light, which forms one of the two foundation pillars on which the theory is based, would be *refuted.* There is, however, in my opinion, *practically no likelihood* that Mr. Miller is right".

Hentschel writes: "Einstein vigorously insisted on the validity of his theories in their original form. Some Einstein articles and interviews in the daily press of the mid – 20's illustrate that he *was not willing to modify his theories but rather was prepared to give them up completely in the case of irrefutable contrary empirical evidence*". Hentschel asks, "Why did he make these risky, but as we all know, well-placed bets on the unaltered theory instead of searching for small modifications, as many of his

---

contemporary colleagues did whenever experiments appeared to come in conflict with his predictions?" [107]

The answer is: Since Special Relativity is a heuristic system of two principles, and it is not a constructive theory like the ether-based electron theory, then one cannot modify principles without giving up the whole theory. However, a theory of principle has a solid theoretical basis, and therefore there is little chance that experiments like that of Miller (and also like that of Walter Kaufmann) would turn to be right.

**4.6.2 Miller's Experiments**

Miller performed his first observations during April 8-21, 1921 when Einstein first visited America with Chaim Weizmann, president of the World Zionist Organization. They both raised funds for the Hebrew University of Jerusalem. In May 1921, Einstein delivered four lectures on relativity theory at Princeton University and received an honorary degree. While he was there, word reached Princeton that Miller had found a nonzero ether drift during preliminary experiments performed in April at Mount Wilson Observatory. Upon hearing this rumor, Einstein produced one of his classical *aperçus*: "The Lord God is Subtle, but malicious he is not" ("Raffiniert ist der Herrgott, aber boshaft ist er nicht"). Nevertheless, on May 25, 1921, shortly before his departure from the United States, Einstein paid a visit to Miller in Cleveland, as discussed below.[108]

In March 1921, Miller set up his interferometer on Mount Wilson, "on the grounds of the Mount Wilson Observatory, on Rock Crusher Knoll or 'Ether Rock' as it came to be called".[109] Miller made numerous observations during the period April 8-21, 1921. The data indicated a possible small periodic effect, with an average second harmonic amplitude of about 0.04 fringe: "The first observations of sixty-seven sets consisting of 350 turns gave a positive effect such as would be produced by a real ether-drift, corresponding to a relative motion of the earth and ether of about ten kilometers per second".[110]

However, Miller suspected that temperature effects or magnetostriction in the steel base of the interferometer as it rotated in the earth's magnetic field might be the cause for this effect: "Before announcing such a result, it seemed necessary to study every possible cause which might produce a displacement of fringes similar to that caused by ether-drift". And so "In order to test the first" Miller covered the metal parts with cork, performed sets of observations, and the data indicated as before a small periodic effect; therefore Miller concluded that radiant heat is not the cause of the observed effect. He continued to perform ether-drift experiments until 1925-1926.[111]

---

[107] Hentschel, 1992, p. 594.
[108] Pais, 1982, p. 113. And this is the title of Abraham Pais' book of 1982.
[109] Miller, Dayton, Clarence, "The Ether-Drift Experiments and the Determination of the Absolute Motion of the Earth", *Reviews of Modern Physics* 5, 1933, pp. 203-242; pp. 217-218.
[110] Miller, 1933, p. 218.

Shankland said, "Undoubtedly the greatest incentive to continue the experiments came from Professor Albert Einstein who visited Miller at Case on May 25, 1921, and urged that further trials be made to remove any possible doubts concerning the earlier results obtained in this experiment".[112] On April 14, 1921, *a month before* Einstein had visited his laboratory, Miller wrote in his notebook: "Sun is shining full on side of house. There was a very large drift which seems to be in the direction of the sun; indicating possibility that the entire effect is due to temperature!"[113]

In his meeting with Einstein Miller probably expressed his concerns about the effect due to temperature. Einstein was very likely convinced that a temperature difference could certainly explain Miller's result: decrease in temperature (i.e., thermal effects) led to change of the optical paths lengths of the interferometer arms in Miller's experiment and this led to the shift in the position of the interference fringes.

Between Shankland's first two conversations with Einstein in 1950, Shankland along with other Case researchers planned to reanalyze Miller's observational data. In April 1955, the very month in which Einstein passed away, the *Reviews of Modern Physics* published a paper on Miller's experiments by Shankland with three colleagues from Case institute of physics. In this paper Shankland summarized a 1954 research, which checked a proposal that could have sprang from Einstein. In their 1950 conversations, Shankland possibly heard from Einstein his opinion that the effect was due to temperature.

---

[111] Miller, 1933, p. 218.
Miller suspected that thermal effects might be important, but he ruled out this possibility. After 1921 "An extended series of experiments was made to determine the influence of inequality of temperature in the interferometer room and of radiant heat falling on the interferometer. Several electric heaters were used, of the type having a heated coil near the focus of a concave reflector. Inequalities in the temperature of the room caused a slow but steady drifting of the fringe system to one side but caused no periodic displacement. Even when two of the heaters, placed at a distance of three feet from the interferometer as it rotated, were adjusted to throw the heat directly on the uncovered steel frame, there was no periodic effect that was measurable. When the heaters were directed to the air in the light-path which had a covering of glass, a periodic effect could be obtained only when the glass was partly covered with opaque material in a very nonsymmetrical manner, as when one arm of the interferometer was completely protected by a covering of corrugated paper-board while the other arms were unprotected. These experiments proved that under the conditions of actual observation, the periodic displacements could not possibly be produced by temperature effects". Miller, 1933, p. 220.
[112] Shankland, McCuskey, Leone, and Kuerti, 1955, pp. 167-168; Pais, 1982, p. 113.
[113] Shankland, McCuskey, Leone, and Kuerti, 1955, p. 174.

Shankland demonstrated that "Miller's extensive Mount Wilson data contain no effect of the kind predicted by the aether theory", because in the laboratory tests the electric heaters were placed at the level of the mirrors and about three feet from the circle travelled by them. The altered refractive index of the heated air and the thermal effects on the mirror supports change of the optical path lengths of the interferometer, and when these are affected unequally in the two arms, the fringes shift in position. On the assumption that the four arms of the interferometer all have the same thermal insulation, a localized temperature gradient across the room will produce the required effect as the instrument rotates, similar to that anticipated for an ether drift.[114]

On August 31, 1954, Einstein sent Shankland a letter expressing his appreciation for his 1954 work:[115]

"You have shown convincingly that the observed effect is outside the range of accidental deviations and must, therefore, have a systematic cause. You made it quite probable that this systematic cause has nothing to do with 'ether-wind', but has to do with differences of temperature of the air traversed by the two light bundles which produced the bands of interference".

## 5 Einstein Route to Special Relativity from 1895 to 1903-1904

Einstein's route to special relativity seems to include the following stages:

**1) In 1894-1895** Einstein wrote an essay and sent it to his uncle Caesar Koch. He then believed in the ether and had heard of Hertz's experiments on the propagation of electromagnetic waves; but he did not show any knowledge of Maxwell's theory. The covering letter that Einstein had sent with the essay to his uncle, received later the added note in Einstein's own hand: '1894 or 95. In 1895 at the age of 16, Einstein was also familiar with the principle of relativity in mechanics.

**2)** A year later, **in 1895-1896**, while in Aarau, Einstein conceived of a thought experiment: what would happen if an observer tried to chase a light wave? Could he catch up with it? If so, he ought to see a standing light wave form, which somehow seemed strange to him.

**3) In 1899** Einstein studied Maxwell's electromagnetic theory, including Hertz's papers on Maxwell's theory. Einstein wrote Marić probably on August 10, 1899, that he is "now rereading Hertz propagation of electric force with great care […] I'm convinced more and more that the electrodynamics of moving bodies as it is presented today doesn't correspond to reality, and that it will be possible to present it in a simpler way". Einstein thought that "electrical forces can be directly defined only for empty space, something also emphasized by Hertz". He reasoned that "Electrodynamics would then be the theory of the movements of moving electricities and magnetisms in empty space.[116] Recall that Reiser wrote, "With a veritable mania for reading, day and night, he went through the works of the great physicists – Kirchhoff, Hertz, Helmholtz, Föppl".[117]

---

[114] Shankland, McCuskey, Leone, and Kuerti, 1955, p. 174.
[115] Lalli, Roberto, "The Reception of Miller's Ether-Drift Experiments in the USA: The History of a Controversy in Relativity Revolution", *Annals of Science* 69, 2012, pp. 153-214, p. 58.

**4) Around 1898-1900** Weber's courses (which Einstein took at the Polytechnic) and Einstein's technical background from childhood (and his above self reading of writings pertaining to electromagnetism and electrodynmics) could have combined to lead him to invent the magnet and conductor thought experiment.

**5) Between 1899 and 1900** Einstein was occupied with the contradiction between the Galilean principle of relativity and the constancy of the velocity of light in Maxwell's theory. Maja wrote in her biography: "Noch wahrend des Studiums an der Hochschule beschäftigen Albert Einstein die Probleme, welche Ausgangspunkte für seine theorien wurden (Widerspruch der Klassischehen Mechaniken und der Konstanz der Lichtgeswindigkeit)".[118] Einstein was troubled by the apparent conflict or discrepancy between classical mechanics and electrodynamics (Maxwell's theory), but was still not acquainted with Lorentz's theory of electrons. Maja then writes that next year (1900) Einstein graduated and got a diploma from the Polytechnic.

**6) Between 1899 and 1901** Einstein was interested in ether drift experiments, and appears to have designed at least two experiments, the first in 1899. Later he was reported to have said: "Then I myself wanted to verify the flow of the ether with respect to the Earth [...]. When I first thought about this problem, I did not doubt the existence of the ether or the motion of the Earth through it".[119] Two years later, after graduating the Polytechnic, Einstein designed a second ether drift experiment. On December 19, 1901 he wrote Marić, "I spent all afternoon with Professor Alfred Kleiner in Zurich telling him my ideas about the electrodynamics of moving bodies, […] He advised me to publish my ideas on the electromagnetic theory of light of moving bodies along with the experimental method".[120]

---

[116] Einstein to Marić, letter 8, August 1899, *CPAE*, Vol 1, Doc. 52, p. 226, Renn and Schulmann, 1992, letter 8; Hertz, Heinrich, "Über die Grundgleichungen der Elektrodynamik für bewegte Körper", *Annalen der Physik und Chemie* 41 (1890), pp. 369-399.
[117] Reiser, 1930, p. 49.
[118] Winteler-Einstein Maja, *Albert Einstein –Beitrag für sein Lebensbild*, 1924, reprinted in abridged form in *The collected papers of Albert Einstein*, Vol 1, 1987, pp xlviii-lxvi. Einstein Archives, Jerusalem: the full biography printed in a typing machine with a few missing pages and double pages, 1924, p. 18.
[119] Einstein, 1922, p. 46.
[120] Einstein to Marić, letter 130, Dec 19 1901, *CPAE*, Vol 1, Doc. 130, p. 328, Renn and Schulmann, 1992, letter 47.

**7) On December 17, 1901** Einstein wrote Marić: "I am busily at work on an electrodynamics of moving bodies, which promises to be quite a capital piece of work. I wrote to you that I doubted the correctness of the ideas about relative motion, but my reservations were based on a simple calculational error. Now I believe in them more than ever".[121] **In 1901** Einstein still accepted *the Galilean kinematics of space and time*, in which the *Galileian principle of relativity* holds good. Einstein realized that – once we consider the induction term in the magnet and conductor thought experiment – the Galilei transformations lead to an asymmetry in the explanation. He felt that there should only be an electric field acting on the electrons in the conductor. In 1901, Einstein still had no solution to the problem of electrodynamics.

**8) In 1902** Einstein read Lorentz's 1895 memoir, *Versuch einer Theorie der elektrischen und optischen Erscheinungen in bewegten Körpern*. **On December 28, 1901,** Einstein wrote Marić, "I want to get down to business now and read what Lorentz and Drude have written about the electrodynamics of moving bodies".[122]

**9)** It appears that **between 1901 and 1903**, Einstein was working with the Maxwell-Hertz equations for empty space. He tried to find solutions to two problems:

**A)** Magnet and conductor experiment and Faraday's induction law lead to conclusion that there is an asymmetry in the explanation depending on whether the magnet moves or the conductor moves. Einstein analyzed the magnet and conductor thought experiment according to Maxwell's theory and the Galilean transformations. But covariance of Maxwell equations failed.

He wrote in 1920, "According to Faraday, during the relative motion of a magnet and an electric circuit, an electric current is induced in the latter. Whether the magnet is moved or the conductor doesn't matter; it [the electric current] only depends on the relative motion". But then **around 1902** when he read Lorentz's *Versuch* he realized, "But according to the Maxwell-Lorentz theory, the theoretical interpretation of the phenomenon is very different for the two cases".[123] Einstein thus chose the relativity principle and dropped Lorentz's theory.

B) *Conflict* between the principle of Galilean relativity and the constancy of the velocity of light.

Max Wertheimer wrote, "In the Maxwell equations of the electromagnetic field, the velocity of light plays an important role; and it is constant. If the Maxwell equations are valid with regard to one system, they are not valid in another. They would have to be changed. When one tries to do so in such a way that the velocity of light is not assumed to be constant, the matter becomes very complicated. For years Einstein tried to clarify the problem by studying and trying to change the Maxwell equations. He did not succeed in formulating these equations in such a way as to meet these difficulties satisfactory. He tried hard to see clearly the relation between the velocity of light and the facts of movement in mechanics. But in whatever way he tried to unify the question of mechanical movement with the electromagnetic phenomena, he got into difficulties". [124]

---

[121] Einstein to Marić, letter 128, Dec 17 1901, *CPAE*, Vol 1, Doc. 128, p. 325, Renn and Schulmann, 1992, letter 46.
[122] Einstein to Marić, letter 131, Dec 28 1901, *CPAE*, Vol 1, Doc. 130, p. 328, Renn and Schulmann, 1992, letter 48.
[123] Einstein, 1920, *CPAE*, Vol 7, Doc. 31, p. 20.

**10) Between 1901 and 1903**, Einstein dropped the ether hypothesis and chose the principle of relativity instead of the postulate of the constancy of the velocity of light, and found a (temporary) solution for his conflict in the form of an emission theory.

Wertheimer wrote, "One of his questions was: What would happen to the Maxwell equations and to their agreement with the facts if one were to assume that the velocity of light depends on the motion of the source of the light?" [125] Einstein replaced Lorentz's theory by emission theory. What led Einstein to the realization that, the ether was superfluous and space is empty (of ether)? Emission theories probably encouraged Einstein to think that the ether is superfluous.

**11)** Einstein seemed to have pondered with this problem for an extra year, **from 1903-1904 until almost spring-summer 1904**.

**12)** According to the 1922 Kyoto lecture notes Einstein said,[126]

"At that time I firmly believed that the electrodynamic equations of Maxwell and Lorentz were correct. Furthermore, the assumption that these equations should hold in the reference frame of the moving body leads to the concept of the invariance of the velocity of light, which however contradicts the addition rule of velocities used in mechanics.

Why do these two concepts contradict each other? I realized that this difficulty was really hard to resolve. I spent almost a year in vain trying to modify the idea of Lorentz in the hope of resolving this problem".

---

[124] Wertheimer, 1916/1945, p. 171.
Einstein also tried to discuss Fizeau's experiment in Lorentz's theory. In the *Versuch* Lorentz managed to derive the Fresnel Formula from the first principles of his theory (stationary ether and moving electrons) without the need of any partial ether drag. Lorentz thus adhered to Fizeau's original 1851 experimental result, but *not* to Fresnel's theoretical interpretation of partial ether drag hypothesis, used to derive his dragging coefficient.
According to the 1922 Kyoto lecture notes Einstein said: "Then I tried to discuss the Fizeau experiment on the assumption that the Lorentz equations for electrons should hold in the frame of reference of the moving body as well as in the frame of reference of the vacuum as originally discussed by Lorentz". Einstein, 1922, p. 46.
[125] Wertheimer, 1916/1945, p. 171.
[126] Einstein, 1922, p. 46.

Hence, **towards spring-summer 1904** Einstein dropped emission theory and returned to Lorentz's theory. He spent almost a year in vain trying to modify the idea of Lorentz in the hope of resolving the above problem until he found the final solution, "the step", **in spring 1905**, which solved his dilemma.

## 5 Emission theory and ether drift experiments

Emission theorists suggested that, what seemed to them as "all the apparent paradoxes of Einstein's theory" might be avoided and at the same time the principle of the relativity of motion retained, if an alternative postulate were true; namely, that the speed of light depends on the motion of its source. In other words, light from a source moving relative to an observer has a velocity equal to the vector sum of the velocity of light from a stationary source and the velocity relative to the observer of the source itself at the instant of emission. Relativity theories based on the principle of relativity and such a postulate were called "emission theories".[127]

**5.1 Ritz's Emission Theory**

In 1912 Paul Ehrenfest published a paper comparing Einstein's views on light propagation with those of Walter Ritz.[128] Ehrenfest noted that although both approaches involved a particulate description of light – Einstein invoked the quanta of light and Ritz proposed particles of light – nevertheless, Ritz's theory constituted a "real" emission theory (in the Newtonian sense), while Einstein's theory was more akin to the ether conception; since it postulated that the velocity of light is independent of the velocity of its source. Ehrenfest suggested possible experiments to distinguish between the two theories and noted the necessary of carrying out such empirical test.

On April 25, 1912, Einstein reacted to Ehrenfest's above paper by writing him the following,[129]

"I was not annoyed in the least by your article! On the contrary, such considerations are quite familiar to me from the pre-relativistic time. I certainly knew that the principle of the constancy of the velocity of light is completely independent of the relativity postulate; and I considered what would be more probable, the principle of the constancy of $c$, as was demanded by Maxwell's equations, or the constancy of $c$, exclusively for an observer sitting at the light source. I decided in favor of the first, since I was convinced that every light should be defined by frequency and intensity alone, completely independently of whether it comes from a moving or resting light source. Moreover it did not occur to me to consider whether the radiation deflected at a point could behave differently in propagation compared to newly emitted radiation from the point concerned. Such complications seemed to me much more unwarranted than those brought by the new concept of time".

---

[127] Tolman, Richard, C., "Some Emission Theories of Light", *The Physical Review* 35, 1912, pp. 136-143, p. 136.
[128] Einstein, 1905c; Ehrenfest, Paul, "Zur Frage der Entbehrlichkeit der Lichtäthers", *Physikalische Zeitschrift* 13 (1912), pp. 317-319.
[129] Einstein to Ehernfest, *CPAE*, Vol 5, Doc. 409.

In his 1912 manuscript on the special theory of relativity, Einstein discusses emission theory and mentions the Swiss theoretical physicist Walter Ritz and Ehrenfest: "In case II the velocity of light in *M* depends on the velocity of motion of the light source with respect to *M* (Ritz and Ehrenfest)".[130]

From what Einstein wrote in his letter to Ehrenfest and in the 1912 manuscript, we can infer that prior to 1905 he seriously considered an emission theory similar to that of Walter Ritz and Ehrenfest's treatment of Ritz's theory.

Ritz[131] first published his emission theory in 1908 with the aim of replacing Einstein's special theory of relativity. Einstein's special theory of relativity assumes as its second postulate the famous light postulate: the velocity of light is independent of the state of motion of its source. After 1905 some scientists felt they could explain the phenomena of electrodynamics without this postulate.

Ritz outlined an emission theory of light that was consistent with classical mechanics in an attempt to develop a new electrodynamics of moving bodies. While Einstein's final solution posited that all light rays travel with the same speed in empty space, Ritz argued that their speeds vary depending on the motion of the sources at the instant of emission – as with any other mechanical projectile. He considered light to consist of particles, and to be shot like projectiles, and not like rays or waves.[132] According to Ritz electrodynamics was thus based on forces and not on fields. If this was the sort of mechanistic picture that Ritz was intending to offer, then it was obvious that Maxwell's field equations were inadequate to describe exactly the laws of propagation of physical actions; first these equations dealt with fields and second,

---

[130] Einstein, Albert, *Einstein's 1912 Manuscript on the Special Theory of Relativity*, 1996, New-York: George Braziller Publishers in association with Jacob E. Safra Philanthropic Foundation and the Israel Museum, Jerusalem, p. 21 (full paragraph is cited further below).
[131] Ritz had done his doctoral thesis on theoretical spectroscopy (atomic spectra) under Woldemar Voigt's supervision in Göttingen. In his dissertation he combined electrodynamics and mechanics to ascertain the laws of spectral series, the work for which he is most well known. Ritz, Walter, "Recherchés critiques sur l'Électrodynamique Gènérale", *Annales des Chimie et de Physique* 13, 2008, pp. 145-275; Ritz, Walter, "Recherchés critiques sur les Théories Électrodynamique des cl. Maxwell et de H. A. Lorentz", *Archives des Sciences Physique et Naturelles* 26, 2008, pp. 209-239.
[132] Martínez, Alberto, "Ritz, Einstein, and the Emission Hypothesis", *Physics in Perspective* 6, 2004, pp. 4-28; p. 4.

Ritz could not accept propagation of waves in medium, such as was the ether.[133] Towards 1908 Ritz became increasingly antagonistic to Maxwell's theory in general and to Lorentz's electron theory in particular. He then sought to devise a synthesis of optics with a new electrodynamics that would account better for experimental facts.[134]

In 1908 Ritz thus objected the ether of electrodynamics theory, and he thought it should be eliminated. "It is clear that one would then has to abandon not only the idea of the existence of an ether but also Maxwell's equations for the vacuum", explained Wolfgang Pauli in his 1921 *Encyclopedia* article on Relativity, "so that the whole of electrodynamics would have to be constructed anew".[135]

Ritz thus thought that, in renouncing classical mechanics, Einstein had paid too high a price to resolve the difficulties at issue. Einstein had changed kinematics, but left the fundamentally-problematic equations of electromagnetism untouched; Einstein assumed that the Maxwell-Lorentz equations were fundamental, and changed Newtonian kinematics. Ritz attempted to solve the problems by taking the inverse route: Let us drop the fundamental field equations of Maxwell-Lorentz theory but leave intact Newtonian kinematics and dynamics, with its force concept and ballistic theory of light, and of course the Galileian principle of relativity.[136]

In 1909 Ritz and Einstein exchanged views on the question of radiation in the *Physikalische Zeitschrift*.[529] Shortly afterwards Ritz died on July 7, 1909, before completing his theory. Two months later, on September 21, Einstein was invited to give his first lecture at the meeting of the *Gesellschaft Deutscher Naturforscher und Ärzte*. Einstein spoke on "The Present Status of the Radiation Problem", and asserted quite at the beginning of the lecture, "It is therefore my opinion that the next stage in the development of theoretical physics will bring as a theory of light that can be understood as a kind of fusion of the wave and emission theories of light".[137]

In 1910 Richard Tolman also suggested an emission theory and in 1912 called Ritz's emission theory, "Ritz's theory of relativity".[138] (Ritz called his theory

---

[133] Martínez, 2004, p. 8.
[134] Martínez, 2004, pp. 6-7.
[135] "Es ist klar, daß damit nicht nur die Existenz des Äthers, sondern auch die *Maxwellschen* Gleichungen für das Vakuum verworfen warden, so daß die ganze Elektrodynamik neu aufgebaut warden muß. Dies hat in einer *systematischen* Theorie nur W. Ritz geten". Pauli, Wolfgang, *Theory of Relativity*, 1958, Oxford and New York: Pergamon, p. 6; Pauli, Wolfgang, "Relativitätstheorie" in *Encyklopädie der Mathematischen wissenschaften*, Vol. 5, Part 2, 1921, Leipzig: Teubner, p. 539, p. 20 (p. 550).
[136] Martínez, 2004, p. 8.
[137] Einstein, Albert, "Über die Entwicklung unserer Anschauungen über das Wesen und die Konstitution der Strahlung", *Deutsche Physikalische Gesellschaft, Verhandlungen* 7, 1909, pp. 482-500; p. 482; *CPAE*, Vol. 2, Doc. 60.
[138] Tolman, 1912, p.141.

"Relativtheorie").[139] Tolman explained that according to Ritz's theory, light retains the component of velocity *v* which it obtained from its original moving source. Thus, all the phenomena of optics would occur as though light were propagated in an ether that is stationary with respect to the original source. Light coming from a terrestrial source would behave as though propagated by an ether stationary with respect to the earth, while light coming from the sun would behave as though propagated by an ether stationary with respect to the sun.[140]

Hence, said Tolman in 1912, "if the Ritz theory should be true, *using the sun as source of light* we should find on rotating the apparatus a shift in the fringes of the same magnitude as originally predicted for the Michelson-Morley apparatus where a terrestrial source was used. If the Einstein theory should be true, we should find no shift in the fringes using any source of light".[141] Tolman suggested the Michelson-Morley experiment would serve as "*experimentum crucis*". Of course, on the basis of either Einstein's or Ritz's theory, there was no ether.

The Michelson experiment with extra-terrestrial light (sun and stars) was actually carried out with negative result by Rudolf Tomaschek (from Heidelberg).[142]

**5.2 Why Einstein Finally Rejected Ritz's Emission Theory**

Einstein told Ehrenfest that before 1905 he himself explored emission theory, before returning to Lorentz's theory; but had abandoned it due to mounting difficulties. He explained to Ehrenfest why an emission theory based on Newton-Galileian kinematics was inadmissible. It led to paradoxical conclusions when attempting to explain such simple things as the reflection of light from a moving mirror.

In his 1912 manuscript on the special theory of relativity Einstein added a crossed-out paragraph containing objections to the explanation of Fizeau's experiment on the basis of emission theory:[143]

"In case II the velocity of light in *M* depends on the velocity of motion of the light source with respect to *M* (~~Ritz and Ehrenfest~~). This being so, light rays of all possible propagation velocities, arbitrary small or arbitrary large, could occur in M. Intensity, color, and Polarization state would not suffice to define a plane light wave; one would have also to add the determinative element of *velocity* […] one is forced to make the most peculiar assumptions if one pursues this point of view, as for example the following: if light of velocity $c + v$ strikes a mirror perpendicularly, then the reflected light has the velocity $c - v$. These complications make it seem understandable why it has not proved possible so far to set up differential equations and boundary conditions that would do justice to this conception".

---

[139] Martínez, 2004, p. 9.
[140] Tolman, 1912, p.141.
[141] Tolman,1912, p.143.
[142] Pauli, 1958, pp. 8; Pauli, 1921/2000, pp. 22-23 (pp. 552-553); Tomaschek , R., "The conduct of light of extraterrestrial light sources", *Annalen der Physik* 73 (1924), pp.105–126.
[143] Einstein, 1912, p. 21.

In 1950, Shankland came to talk with Einstein on Ritz's theory because of his personal interest in the Michelson-Morley experiment and Miller's repetitions of it. Shankland cited the experiments which had disproved Ritz's emission theory:[144]

"I mentioned the experiments which disproved the Ritz emission theory of light especially de Sitter's work on spectroscopic binaries and the null result obtained by D.C. Miller [in 1925] at Case with the modification of the Michelson-Morley experiment using sunlight. He [Einstein] said he knew de Sitter well but told me he considered the most decisive experiment along these lines to be the repetition of the Michelson-Morley experiment performed with starlight at Heidelberg by a student of Lenard's (Tomaschek) [in 1924] […] Lenard, who along with Stark, was the most violently Nazi of all German scientists was referred to by Einstein with complete fairness and with not a slightest trace of malice or bitterness.

This led him to a discussion of emission theories of light, and he told me that he had thought of, and abandoned the (Ritz) emission theory before 1905. He gave up this approach because he could think of no form of differential equations which could have solutions representing waves whose velocity depended on the motion of the source. In this case, the emission theory would lead to phase relations such that the propagated light would be all badly 'mixed up' and might even 'back up on itself'. He asked me, 'Do you understand that?' I said no and he carefully repeated it all. When he came again to the 'mixed up' part he waved his hands before his face and laughed, an open heartily laugh at the idea!

Then he continued. 'The theoretical possibilities in a given case are relatively few and relatively simple, and among them the choice can often be made by quite general arguments. Considering these tells us what is possible but does not tell us what reality is'.

When I suggested that Ritz's theory was the best of the several emission theories of light, he shook his head and replied that Ritz's theory is very bad in spots. But he quickly added, 'Ritz made a great contribution when he showed that frequency differences are the crucial thing in spectral series' ".

On February 1952, Einstein elaborated on this point:[145]

"Before setting up the special theory of rel., I had myself thought of investigating such a possibility [emission theory]. At that time I had only a weighing of the plausibility of theoretical arguments at my disposal. […] I deliberated as follows: If a suitably accelerated light source emits light in one direction (e.g., the direction of the acceleration), then the planes of equal phase move with different speed, […] so all the surfaces of equal phase coincide at a particular place, so that the wavelength there is infinitely small. Moreover the light will be so turned around that the rear part overtakes the front".

---

[144] Shankland, 1963, p. 49.
[145] Quoted in Norton, 2004, p. 72.

Einstein despaired of finding the answer to all his problems in electrodynamics of moving bodies and the structure of matter and radiation by an emission theory. He decided to come back to the Maxwell-Lorentz theory, and to reconsider it again. He was looking for a way that after all would make the Maxwell-Lorentz equations compatible with the principle of relativity once one abandoned the ether concept.

## 6 "The Step"

Pais wrote, "When I talked with him about those times of transition, he expressed himself in a curiously impersonal way. He would refer to the birth of special relativity as 'den Schritt' the step".[582] When was the special relativity conceived? In 1895 when Einstein invented the chasing after a light beam thought experiment? In 1900 when he was troubled by the conflict between classical mechanics and the constancy of the velocity of light? When he propounded the magnet and conductor thought experiment? When he dropped emission theories and came back to Lorentz's theory?

In his *Autobiographical Notes* Einstein discussed his objections to Newton's theory while he was trying to develop his General Theory of Relativity. Then suddenly he addressed Newton directly in one of the most brilliant paragraphs of the *Notes*:[146]

"Enough of this. Newton, forgive me; you found the only way which, in your age, was just about possible for a man of highest thought – and creative power. The concepts, which you created, are even today still guiding our thinking in physics, although we now know that they will have to be replaced by others farther removed from the sphere of immediate experience, if we aim at a profounder understanding of relationships."

Einstein was struggling emotionally twice when he replaced Newton's concepts: once before 1905, and a second time when he worked on the General Theory of Relativity. Before 1905 it took Einstein almost five years to replace Newton's concepts. He tried to replace Lorentz first, and leave Newton intact. But finally Einstein realized that he must replace Newtonian kinematics with another kinematics in order to arrive at a profounder understanding of the electrodynamics and optics of moving bodies. As said before, little is known of Einstein's struggles between 1902 and 1905, but in the previous sections I tried to reconstruct what Einstein perhaps could have done in these years. Evidence is nevertheless available about Einstein's final step from **May 1905 to**

---

[146] Einstein, 1949, pp. 30-33.

**June 1905**, the breakthrough which led him to the theory of special relativity. According to the available evidence it appears that, within the five to six weeks after May 1905, Einstein arrived at the solution to his problem above, and was able to complete and submit his paper "Electrodynamics of moving Bodies" for publication.

According to Stachel, Einstein confronted a new problem after he gave up the ether. According to the Maxwell-Hertz equations, there exists one inertial reference frame in which the speed of light is constant regardless of the motion of the light source. If there is no ether to single out this inertial frame, then Einstein's principle of relativity requires that it must hold in all inertial frames.

Einstein realized that the ordinary Newtonian law of addition of velocities was probably problematic ("*Newton verzeih' mir*"): In all of his struggles with an emission theory as well as with Lorentz's theory, Einstein had been assuming that the ordinary Newtonian law of addition of velocities was unproblematic. However, if the Newtonian addition law of velocities was correct then there would only be one inertial frame in which the velocity of light was constant and independent of direction. This violated the principle of relativity and supplied the reason why Einstein had abandoned Lorentz's theory in the first place and resorted to an emission theory of light.

This law of Newtonian mechanics was based on certain tacit assumptions made about the nature of time, simultaneity, and space measurements. In order to solve the contradictions, Einstein had to abandon the Newtonian law of addition of velocities, and with it the tacit assumptions implicit in this law. He accepted a new law of addition of velocities, and in so doing he needed to make new kinematical tacit assumptions about space and time.

Einstein defined distant simultaneity physically, and constructed a new kinematical theory based on the relativity principle and the light principle, thus resolving the apparent conflict between them.[147]

In making Lorentz's light postulate compatible with the principle of relativity, Einstein very likely realized the reciprocity of the appearances, which does not presuppose the new concept of time.[148]

---

[147] Satchel, 1989a, in Stachel, 2002, pp. 197, 199, 200. Stachel, 1989b, in Stachel 2002, pp. 164, 165-166.

[148] Einstein held in his personal library two copies of *Gulliver's Travels*, one in German and the other in English. Einstein Archives, Jerusalem. Einstein could have been inspired to think about the reciprocity of appearances by books that he liked, and later he owned in his personal library in Princeton, such as *Gulliver Travels*. Arthur Eddington wrote in his book, *Space, Time & Gravitation*, "it is the reciprocity of these appearances – that each party should think the other has contracted – that is so difficult to realize. Here is a paradox beyond even the imagination of Dean Swift. Gulliver regarded the Lilliputians as a race of dwarfs; and the Lilliputians regarded

# 7 Einstein's steps toward the "The Step"

## 7.1 Five to six weeks between the discovery and the relativity paper

On **May 18 or 25, 1905** Einstein wrote the famous letter to his friend Habicht telling him about the four or five path breaking papers he was producing, "The fourth paper is only a rough draft at this point, and is an electrodynamics of moving bodies which employs a modification of the theory of space and time; the purely kinematical part of this paper will surely interest you".[149]

The *Annalen der Physik* received the paper on **June 30, 1905**. Hence 5 to 6 weeks before completing his relativity paper, Einstein had only "a rough draft" of it.[150]

This draft presented a modification of the theory of space and time – very likely the physical definition of simultaneity had already been formulated, and the draft had a purely kinematical part.

In 1952 Seelig asked Einstein: when was the birth of special relativity? Einstein replied on March 11, 1952 the following answer,[151]

"Zwischen der Konzeption der idee der speziellen Relativitätstheorie und der Beendigung der betreffenden Publikation sind fünf oder sechs Wochen vergangen. Es würde aber kaum berechtigt sein, dieses als Geburtstag zu bezeichnen, nachdem doch

---

Gulliver as a giant. That is natural. If the Lilliputians had appeared dwarfs to Gulliver, and Gulliver had appeared a dwarf to the Lilliputians – but no! that is too absurd for fiction, and is an idea only to be found in the sober pages of science". Eddington, Sir Arthur, *Space, Time & Gravitation*, 1920/1995, Cambridge: Cambridge, pp. 23-24.
[149] Einstein to Habicht, 18 or 25 May, 1905, *CPAE*, Vol. 5, Doc. 27.
[150] Reiser wrote in his biography of Einstein, "He had made the revolutionary discovery that the traditional conception of the absolute character of simultaneity was a mere prejudice, and that the velocity of light was independent of the motion of coordinate systems. Only five weeks elapsed between this discovery and the first formulation of the special theory of relativity in the treatise entitled 'Towards the Electrodynamics of Moving Bodies' (published in 1905)". Reiser, 1930, p. 69. In Wertheimer's book we find similar remarks: "It was then that the great problem really started to trouble" Einstein. "He was intensely concerned with it for seven years; from the moment, however, that he came to question the customary concept of time […], it took him only five weeks to write his paper on relativity," (Wertheimer, 1916/1945, p. 169) and, "I wish to report some characteristic remarks of Einstein himself. Before the discovery that the crucial point, the solution, laid in the concept of time, more particularly in that of simultaneity, axioms played no role in the thought process – of this Einstein is sure. (The very moment he saw the gap, and realized the relevance of simultaneity, he knew this to be the crucial point for the solution.) But even afterward, in the final five weeks, 'No really productive man thinks in such a paper fashion', said Einstein". Wertheimer, 1916/1945, p. 183, note. 7.
[151] Einstein to Seelig, March 11, 1952, *ETH-Bibliothek, Zürich - Archive und Nachlässe*.

vorher die Argumente und Bausteine jahrelang vorbereitet worden waren, allerdings ohne die endgiltige Entscheidung vorher zu bringen".

In his book Seelig reported Einstein answer: "To my question, whether the birth of the relativity theory could be pinpointed to a certain year, as in the case of Planck's quantum theory, Professor Einstein replied to me on March 11[th] 1952: 'Five or six weeks elapsed between the completion of the conception of the idea of the special relativity theory and the completion of the appropriate publication. This, however, can hardly be considered as a birthday, since previously the arguments and the foundation stones had been prepared over a period of many years, though without bringing the ultimate decision.'".[152]

**7.2 The Einstein-Besso meeting**

According to the Kyoto lecture notes Einstein described the final stages of his work on the theory of relativity in the following way,

"By chance a friend of mine in Bern [Besso] helped me out. It was a beautiful day when I visited him with this problem. I started the conversation with him in the following way: 'Recently I have been working on a difficult problem. Today I come here to battle against that problem with you'. We discussed every aspect of this problem. Then suddenly I understood where the key to this problem lay. Next day I came back to him again and said to him, without even saying hello, 'Thank you. I've completely solved the problem'. An analysis of the concept of time was my solution. Time cannot be absolutely defined, and there is an inseparable relation between time and signal velocity. With this new concept, I could resolve all the difficulties completely for the first time.

Within five weeks the special theory of relativity was completed."[153]

The scenario could be the following. Einstein could have visited and consulted his close friend Michele Besso, whom he thanked at the end of his relativity paper, "In conclusion, I note that when I worked on the problem discussed here, my friend and colleague M. Besso faithfully stood by me, and I am indebted to him for several valuable suggestions".[154] What could Besso's valuable suggestions have been?

Toward the end of 1903 a vacancy for a "technical expert II class" examiner in the Patent Office was advertised. Einstein immediately drew Michele Besso's attention and the latter joined the Patent Office. Since **March 15, 1904** Besso worked at the Patent Office in Bern. Einstein and his wife Mileva gave up their apartment at Kramgasse 49 in the old city center on **May 15, 1905** and moved to Besenscheuerweg

---

[152] Seelig, 1956, pp. 68-69; Seelig, 1954, p. 82.
[153] Einstein, 1922, p. 46.
[154] Einstein, 1905a, p. 921.

28 in the Mattenhof district on the outskirts. They moved to be closer to Michele Besso and his wife. The two, Einstein and Besso, therefore could go to and from the Patent Office together.

Besso and Einstein were very close friends; Einstein opened his letter to Besso from June 23, 1918, by saying, "When I see your writing I always have a very special feeling, because nobody is so close to me, nobody knows me so well, nobody is so kindly inclined to me as you are".[155] Besso died a month before Einstein. Einstein wrote to Besso's family, "Later, the Patent Office brought us together again. The conversations on the way home were of incomparable charm – it was as if the 'All-too-human' concerns did not at all exist".[156]

Seelig writes: "the first friend to hear of the relativity theory was the engineer, Michele Angelo Besso.[…] Since the two civil servants went the same home and Besso was always eager to discuss the subjects of which he knew a great deal – sociology, medicine, mathematics, physics and philosophy – Einstein initiated him into his discovery. Besso immediately recognized it as a discovery of the utmost importance and of the greatest consequence. The main subject of discussion was the discovery of the light quanta. In endless conversations his cultured friend, in the role of a critical disbeliever, defended Newton's recognized time and space concepts, into which he wove Mach's sensualistic positivism, and his analytical criticism of Newtonian mechanics.[157] 'I could not have found a better sounding-board in the whole of Europe', Einstein remarked when the conversation turned one day to Besso".[158]

From Seelig's above report, "In endless conversations" Besso, "in the role of a critical disbeliever, defended Newton's recognized time and space concepts". But that didn't bother Einstein, because, Besso was the best "sounding-board in the whole of Europe". On the contrary, Einstein needed someone to defend the conventional Newtonian time and space concepts; only this way he could realize that these concepts were to be replaced by new ones.

---

[155] Einstein to Besso, June 23, 1918, Einstein and Besso, Spezieli, 1971, letter 43.
[156] Einstein to Besso's family, March 21, 1955, Einstein and Besso, Spezieli, 1971, letter 215.
[157] While a student at the Polytechnic, Einstein read two of Ernst Mach's historical-critical studies, the 1897 *Die Mechanik in ihrer Entwicklung/Historisch-Kritisch dargestellt* (The Science of Mechanics), and the 1896 *Die Prinzipien der Wärmelehre/Historisch-Kritisch Entwickelt* (Principles of the Theory of Heat), Einstein to Marić, 10 Sept, 1899, *CPAE*, vol 1, Doc. 54; Renn and Schulmann, 1992, letter 10. Michele Besso recommended these to Einstein in 1897. Besso asked Einstein in 1948 about Mach's influence on his thought. Einstein replied to Besso from Princeton on 6 January 1948 a long explanation on Mach's influence on his thought. Einstein to Besso, January 6, 1948, Einstein and Besso, Speziali, 1971, Letter 153.
[158] Seelig, 1956, p. 71; Seelig, 1954, p. 85.

## 7.3 The final discovery within five weeks

Let us try to find out what could be Einstein's final steps *after* he had arrived at "The step". Einstein was about to formulate a draft of the kinematical part of his relativity paper;[159] Einstein explained the main idea of this section to his colleagues from the patent office, especially to Joseph Sauter.

Seelig wrote that as to the electrodynamics of moving bodies, Sauter told him that he could still remember "today" (approximately 1952), how Einstein, in the spring of 1905, explained to him in great excitement his discovery of the relativity theory as they walked home together. He gave him his notes, which Sauter criticized with official severity; "I pestered him for a whole month with every possible objection without managing to make him in the least impatient, until I was finally convinced that my objections were no more than the usual judgments of contemporary physics".[160]

Einstein explained to Sauter his discovery in great excitement, and Einstein was patient. Einstein perhaps explained to Sauter his discovery **at the beginning of May 1905** (approximately at the time he wrote Habicht) and he gave Sauter his notes (a rough draft of the kinematical part of the relativity paper) around this time. Sauter pestered Einstein for a whole month until perhaps **the beginning of June**.

Fifty years later Sauter wrote in his memories "Comment j'ai appris à connaître Einstein":[161]

"Before any other theoretical consideration, Einstein pointed out the necessity of a new definition of 'synchronization' of two identical clocks distant from one another; to fix these ideas, he told me, 'suppose one of the clocks is on a tower at Bern and the other on a tower at Muri (the ancient aristocratic annex of Bern). At the instant when the clock of Bern marks noon exactly, let a luminous signal leave from Bern in the direction of Muri; it will arrive at Muri when the clock at Muri marks a time noon + *t*; at that moment, reflected the signal in the direction of Bern; if on the moment when it returns to Bern the clock in Bern marks noon + *2t*, we will say that the two clocks are synchronized'. After having defined what he meant by simultaneity of two instantaneous events produced in 2 distant points, Einstein defined the two principles on which he based all the calculations of his new physics".

But in the same book at which Sauter's memoirs are found, Flückiger's *Einstein in Bern*, Flückiger writes, that the first friend who became acquainted with the theory of relativity in Bern was Einstein's colleague Michele Besso at the Patent Office. The other two friends from the patent office, Dr. Sauter and Lucian Chavan were also introduced to the revolutionary discovery. Flückinger then writes that Sauter remembers well his joint walks with Einstein, and how excited Einstein was with the new discovery.[162] As said above, Seelig wrote that, Einstein explained to Sauter his discovery and gave him notes.[163] These are presumably rough drafts of the kinematical part of the relativity paper.

---

[159] Einstein to Habicht, 18 or 25 May, 1905, *CPAE*, Vol 5, Doc. 27; Seelig, 1956, pp. 74-75; Seelig, 1954, pp. 88-89.
[160] Seelig, 1956, pp. 73-74; Seelig, 1954, pp. 87-88.
[161] Sauter, 1960, in Flückiger, 1960/1974, p. 156.

According to the account of Sauter, "Einstein pointed out the necessity of a new definition of 'synchronization' of two identical clocks distant from one another […]. After having defined what he meant by simultaneity of two instantaneous events produced in 2 distant points, Einstein defined the two principles on which he based all the calculations of his new physics".[164]

Einstein therefore explained to Sauter the contents of the notes he had given him: the necessity of the definition that would appear in section §1 of his paper, the definition for *distant simultaneity*. He explained to him the meaning of this definition. Subsequently he defined the two principles of his new kinematics.

We can therefore guess that by the time that Einstein spoke with Sauter – and explained to him with a thought experiment on two clocks, one on a tower at Bern and the other on a tower at Muri, his definition of distant simultaneity – *he had already written the kinematical part of his 1905 relativity paper*. This happened sometime **in June 1905**.

## 8. P.S.: Did Poincaré have an Effect on Einstein's Pathway toward the Special Theory of Relativity?

On July 16, 1955, at the International Relativity Conference in Bern, Max Born delivered a lecture, "Physics and Relativity", "Physik und Relativität", and spoke about Poincaré's influence on Einstein,

"*Carl Seelig*, who had published a very attractive book '*Einstein* in Switzerland', wrote *Einstein* and asked him which scientific literature had contributed most to his ideas on relativity during his period in Bern and received an answer on February 19[th], 1955, which he published in the *Technische Rundschau* (N. 20, 47, Bern 6, May, 1955).[165]

---

[162] Flückiger, 1960/1974, pp. 102-103.
[163] Seelig, 1956, pp. 73-74; Seelig, 1954, pp. 87-88
[164] Sauter, 1960, in Flückiger, 1960/1974, p. 156.
[165] Born, 1969, pp. 103-104; Born, 1959, pp. 189-190.
*Carl Seelig*, der ein sehr reizvolles Buch "*Einstein* und die Schweiz" veröffentlicht hat, schrieb an *Einstein* die Frage, welche wissenschaftliche Literatur zu seinen Gedanken über Relativität während seiner Berner Periode beigetragen habe, und

Seelig wrote Einstein the following letter,[166]

"Februar 17, 1955

Lieber Herr Professor Einstein,

Erlauben Sie, dass ich eine Ihnen vielleicht einfaeltig erscheinende Frage stelle.

Ich wurde auf das Werk "History of the theories of aether and Electricity" 1900-1926" des Mathematikers Whittaker und im speziellen auf das zweite "Relativity ~~und~~ of Poincaré[167] and Lorentz" aufmerksam gemacht, in dem die merkwuerdige Behauptung steht, dass Poincaré und Lorentz die eigentlichen Begruender der Relat. Theorie seien. Fuer Whittaker scheint das Problem mehr ein mathematischen als ein Physikalisch-philosophsch es zu sein. Es waere sehr gut von Inhen, wenn ich erfahren duerfte, ob Sie insbesondere von Poincaré waehrend Ihrer Berner Zeit, d.h. vor 1905 entscheidende Impulse empfangen haben und wie Whittaker zu dieser behauptung kommen kann?

Vermoegen Sie sich zu erinnern, ob Sie auch schon die Lorentz'schen vor 1905 durchgearbeitet haben? An kommenden Kongress werden solche und aehnliche Fragen auftauchen.

...

Ihr

C.S"

Seelig heard about the work of the mathematician Edmund Whittaker, "History of the Theories of Aether and Electricity 1900-1926", and of the chapter "Relativity of Poincaré and Lorentz". In this chapter, writes Seelig, Whittaker makes the curious claim that Poincaré and Lorentz are the actual founders of the theory of relativity. For Whittaker, says Seelig, the problem seems to be mathematical rather than physical-philosophical.

Seelig requested from Einstein an answer to the question: Whether, as Whittaker claims, before 1905 during Einstein's time in Bern, Poincaré in particular had a decisive impact on him. Seelig also asked Einstein whether he remembered that he worked on Lorentz before 1905. Seelig told the aging Einstein that, in the next congress such questions would be raised. However, Einstein did not live long enough to participate in this conference and answer these questions. Indeed the conference was held in July 1955 (shortly after Einstein died), and Max Born had participated

---

erhielt am 19. Februar 1955 eine Antwort, die er in der Technischen Rundschau (Nr. 20, 47. Jahrgang, Bern 6. Mai 1955) abgedruckt hat.
[166] Item 39 068, Einstein Archives.
[167] "of Poincaré" was written above the word "Relativity" as if Seelig forgot to include Poincaré, and included him later.

and reported about Einstein's reply to Seelig's question in his talk "Physik und Relativität". Einstein answered Seelig's questions in a return letter,[168]

"19.2.55

Lieber Herr Seelig,

Es ist Zweifellos, dass die <u>spezielle</u> Relativitätstheorie, wenn wir ihre Entwicklung rückschauend betrachten, im Jahre 1905 "reif zur Entdeckung war". Lorentz hatte schon erkannt, dass für die Analyse der Maxwell'schen Gleichungen die später nach ihm benannte Transformation wesentlich sei, und Poincaré hat diese Erkenntnis noch vertieft. Was mich betrifft, so kannte ich nur Lorentz' bedeutendes Werk von 1895, aber nicht Lorentz' spätere Arbeit, und auch nicht die daran anschliessende Untersuchung von Poincaré. In diesen Sinne war meine Arbeit von 1905 selbständig. Was dabei neu war, war die Erkenntnis, dass die Bedeutung der Lorentztransformation über den Zusammenhang mit den Maxwell'schen Gleichungen hinausging und das Wesen von Raum und Zeit im Allgemeinen betraf. Auch war die Einsicht neu, dass die "Lorentz-Invarianz" eine allgemeine Bedingung sei für jede physikalische Theorie. Dies war für mich von besonderer Wichtigkeit, weil ich schon früher erkannt hatte, dass die Maxwell'sche Theorie die Mikro-Struktur der Strahlung nicht darstelle und deshalb nicht allgemein haltbar sei.

[...]

Mit besten Grüssen und Wünschen

Ihr A.E"

Born reported about Einstein's *published* reply: [169]

"There is no doubt, that the special theory of relativity, if we regard its development in retrospect, was ripe for discovery in 1905. *Lorentz* had already observed that for the analysis of *Maxwell*'s equations the transformations which later were known by his name are essential, and *Poincaré* had even penetrated deeper into these connections. Concerning myself, I knew only *Lorentz*'s important work of 1895 – 'La théorie électromagné de *Maxwell*' and 'Versuch einer Theorie der elektrischen und Optischen Erscheinungenin bewegten Körpern' – but not *Lorentz*'s later work, nor the consecutive investigations by *Poincaré*. In this sense my work of 1905 was independent. The new feature of it was the realization of the fact that the bearing of the *Lorentz* transformation transcended its connection with *Maxwell*'s equations and was concerned with the nature of space and time in general. A further new result was that the '*Lorentz* invariance' is a general condition for any theory. This was for me of particular importance because I had already previously recognized that *Maxwell*'s theory did not represent the microstructure of radiation and could therefore have no general validity".

---

[168] Item 39 070, Einstein Archives.
[169] Born, 1969, pp. 103-104; Born, 1959, pp. 189-190.

This is the published reply as it appeared in Born's book without a line under the word spezielle".[170]

In his letter to Seelig Einstein put a line under the word "special". Einstein told Seelig that, when developing the special theory of relativity he did not know of Poincaré's studies pertaining to the dynamics of the electron.

In his talk "Geometry and Experience" from 1921, Einstein acknowledged Poincaré's influence on him: Before 1905 Einstein read Poincaré's *La science et l'hypothèse*, and a major part of this book is dedicated to geometry. Einstein wrote that, "sub specie", and, in essence, he agreed with Poincaré's conventionalism of geometry.[171] However, he suggested an alternative view,[172]

According to Einstein, the question whether the continuum is Euclidean or rather according to the general Riemannian scheme, or whether its structure is according to a different geometry, is actually *a physical question, which must be answered by experience.* This is not a question which is to be chosen by convention as a matter of convenience. Einstein said that without this alternative view he would not have been able to formulate the theory of general relativity.[173]

In 1922 Einstein visited and lectured in Paris. The astronomer Charles Nordmann reported about Einstein's visit to Paris, the discussions and meetings. He published his

---

[170] <<*Einstein* schrieb:
"Es ist zweifellos, daß die spezielle Relativitätstheorie, wenn wir ihre Entwicklung rückschauend betrachten, im Jahre 1905 reif zur Entdeckung war. *Lorentz* hatte schon erkannt, daß für die Analyse der *Maxwell*schen Gleichungen, die später nach ihm benannte Transformation wesentlich sei, und *Poincaré* hat diese Erkenntnis noch vertieft. Was mich betrifft, so kannte ich nur *Lorentz*' bedeutendes werk von 1895 – 'La théorie électromagnétique de *Maxwell*' und 'Versuch einer Theorie der elektrischen und optischen Erseeinungen in bewegten Körpern' – aber nicht *Lorentz*' spätere Arbeiten und such nicht die daren anschließende Untersuchung von *Poincaré*. In diesem Sinne was meine arbeit von 1905 selbständig.
Was dabei neu war, war die Erkenntis, daß die Bedeutung der Lorentz-Transformation über den Zusammenhang mit den *Maxwell*schen Gleichungen hinausging und das Wesen von Raum un Zeit im allgemeinen betraf. Auch war die Einsicht neu, daß die '*Lorentz*-Invarianz' eine allgemeine Bedingung sei für jede Theorie. Das war für mich von besonderer Wichtigkeit, weil ich schon früher erkannt, daß die *Maxwell*sch Theorie die Mikrostruktur der Strahlung nicht darstelle und deshalb nicht allgemein halbar sei">>.
[171] Einstein, Albert, "Geometrie und Erfahrung, *Preußische Akademie der Wissenschaften* (Berlin). *Sitzungsberichte* I, 1921, pp.123-130; p. 127.
[172] Einstein, 1921, p. 128.
[173] Einstein, 1921, p. 126; Einstein again spoke about "the practically rigid body", even thought he knew it had no meaning in the theory of relativity, and that there was no rigid body in practice.

report as a paper, "Einstein expose et discute sa théorie". During his stay in Paris, Einstein was asked of Poincaré's influence on his way to his theory of general relativity. Nordmann reported in his paper,[174]

"Henri Poincaré died and certainly it would have been something deeply moving to see Einstein discuss this powerful spirit which paved the way on so many points. Would he become a partisan of the General Theory of Relativity? It is probable, it is not absolutely sure. In many pages of studies celebrating the origins and foundations of geometry, Henri Poincaré had come to the conclusion that, if it is not any more ideal, then the Euclidean Geometry is that which corresponds to the nature of the external world and our sensations. On this point, Einstein's ideas were clearly separated from Poincaré's, from the day he foresaw the bending of light in gravity, for which a test was provided lately, and as we know Poincaré did not consider.

This is the keystone of all relativity, the central point that Einstein could infer that the geometry of the world was actually a non-Euclidean geometry. It is difficult to think what Poincaré would have thought about it. Surely in this form or another, it would have been self-consistent, complete relativism, and he would have accepted it with certain sympathy, one that would have allowed him to live without those mystical creatures that inspired him with strange repulsion: Newton's absolute space and absolute time".

And since Poincaré's and Einstein's conceptions were so far removed one from each other, Nordmann concluded by reporting what Einstein had told him,

"Perhaps more than Poincaré, Einstein admits to have been influenced by the famous Viennese physicist Mach".

---

[174] Nordmann, 1922, pp. 142-143.
"Henri Poincaré est mort et certes c'eût été une chose profondément émouvante que de voir discuter avec Einstein ce puissant esprit qui sur tant de points a montré la voie. Eût-il été partisan de la théorie de la Relativité généralisée? C'est probable, ce n'est pas absolument sûr. Étudiant dans maintes pages célèbres les origines et fondements de la géométrie, Henri Poincaré était arrive à cette conclusion que, si élle n'est pas plus vraie idéalement que les autres, le géométrie euclidienne est celle qui correspond à la nature du monde extérieur et de nos sensations. Sur ce point Einstein s'est nettement séparé des idées poincaristes, à dater du jour où il a prévu l'incurvation par la pesanteur des rayons lumineux qui fut vérifiée naguère comme on sait et que Poincaré n'avait pas envisage.
C'est là la clef de voûte de toute la Relativité, le point central don't Einstein a pu déduire que géometrie réelle du monde est effectivement une géométrie non euclidiene. Il est bien difficile de savoir ce qu'en eût pensé Poincaré. Sûrement sous cette forme ou sous une autre; il eût été, logique avec lui-même, relativiste integral et il eût accepté avec une sympathie certain tout ce qui lui eût permis de vivre sans ces creatures mystiques qui lui inspiraient une repulsion singulière: l'espace absolu et le temps absolu de Newton.
Plus peut-être que celle de Poincaré, Einstein avoue avoir subi l'influence du célèbre physician viennois Mach…"

To be continued in a book that I intend to publish…

**Innovation never comes from the established institution**.

Tel-Aviv, Passover, 2012.

*I wish to thank Prof. John Stachel from the Center for Einstein Studies in Boston University for sitting with me for many hours discussing special relativity and its history. Almost every day, John came with notes on my draft manuscript, directed me to books in his Einstein collection, and gave me copies of his papers on Einstein, which I read with great interest.*